\documentclass[12pt]{article}
\pdfoutput=1

\usepackage{amsmath}
\usepackage{amsthm}
\usepackage{amscd}
\usepackage{amsfonts}
\usepackage[psamsfonts]{amssymb}
\usepackage{graphicx}
\usepackage{epsfig}
\usepackage{hyperref}

\makeatletter
\renewcommand\section{\@startsection {section}{1}{\z@}%
                  {-3.5ex \@plus -1ex \@minus -.2ex}%nn
                  {2.3ex \@plus.2ex}%
                  {\normalfont\large\bfseries}}
\renewcommand\subsection{\@startsection{subsection}{2}{\z@}%
                  {-3.25ex\@plus -1ex \@minus -.2ex}%
                  {1.5ex \@plus .2ex}%
                  {\normalfont\bfseries}}
\renewcommand\subsubsection{\@startsection{subsubsection}{3}{\z@}%
                  {-3.25ex\@plus -1ex \@minus -.2ex}%
                  {1.5ex \@plus .2ex}%
                  {\normalfont\itshape}}
\makeatother

\newcommand\beq{\begin{eqnarray}}
\newcommand\eeq{\end{eqnarray}}

\begin{document}

\begin{titlepage}
\begin{flushright}
CQUeST--2012-0550 \\
TAUP-2955/12
\end{flushright}
%\hskip11.6cm{CQUeST--2012-0550} \\
%\hskip11.6cm{TAUP-2955/12}
\vskip1cm

\bigskip\bigskip

\begin{center}
{\Large{\bf Aging Logarithmic Conformal Field Theory \\
\large : a holographic view }}

\bigskip\bigskip
Seungjoon Hyun$^{1}$, Jaehoon Jeong$^{2}$ and
Bom Soo Kim$^{3}$
\bigskip

${}^1${\it {\small Department of Physics, College of Science, Yonsei University, Seoul 120-749, Korea}} \\
${}^2${\it {\small Center for Quantum Spacetime, Sogang University, Seoul 121-742, Korea}}\\
${}^3${\it {\small Raymond and Beverly Sackler School of Physics and Astronomy,}} \\
{\it {\small Tel Aviv University, 69978, Tel Aviv, Israel}}\\

\end{center}

\bigskip

{\small
\centerline{sjhyun@yonsei.ac.kr, ~~jhjeong@sogang.ac.kr, ~~bskim@post.tau.ac.il}
}

\bigskip\bigskip
\begin{abstract}
We consider logarithmic extensions of the correlation and response functions of scalar operators
for the systems with aging as well as Schr\"odinger symmetry. Aging is known to be the
simplest nonequilibrium phenomena, and its physical significances can be understood by
the two-time correlation and response functions.
Their logarithmic part is completely fixed by the bulk geometry in terms of the conformal
weight of the dual operator and the dual particle number.

Motivated by recent experimental realizations of Kardar-Parisi-Zhang universality class in growth phenomena
and its subsequent theoretical extension to aging, we investigate our two-time
correlation functions out of equilibrium, which show several qualitatively
different behaviors depending on the parameters in our theory.
They exhibit either growing or aging, {\it i.e.} power-law decaying, behaviors for the entire range
of our scaling time.
Surprisingly, for some parameter ranges, they exhibit growing at early times as well as aging at later times.

\end{abstract}

\vspace{0.5in}
\end{titlepage}

\tableofcontents

\section{Introduction}

Most of the physical systems are out of equilibrium \cite{HenkelBook1}\cite{HenkelBook2}.
Thus it is very interesting to understand various physical properties of those systems. 
In general, it has been known to be difficult to understand the nonequilibrium
phenomena despite active studies. Still, progresses have been made in the context of the simple
nonequilibrium critical phenomena, where one can use their underlying local scaling invariance,
universality classes and critical exponents to capture universal properties. One simple example is
Kardar-Parisi-Zhang (KPZ) universality class \cite{KPZ}\cite{BSBook}. This universality class is known to describe
many interesting systems in nature, such as paper wetting, epidemics and fluid flow of porous media
and so on \cite{HenkelBook1}\cite{HinrichsenReviewDP}. In the last few years,
this class is experimentally realized in the context of surface growth in turbulent liquid
crystals \cite{TakeuchiPRL}\cite{TakeuchiSciRep}, which provides the experimental evidence of
the dynamical scaling of the KPZ theory in 1+1 dimensions.

Inspired by these developments, the authors of \cite{Henkel:2011NP}
investigate the possibility of aging phenomena, which can arise in the same
KPZ universality class. It is shown that the systems undergo simple aging in the correlation and
response functions, whose dynamical scaling is characterized by some aging exponents.
Moreover, it is claimed that the form of the autoresponse scaling function is well described by
the logarithmic extension of the scaling function with local scale invariance \cite{Henkel:2010hd}.
This provides us the main physical motivation to pursue the logarithmic extension of the aging
correlation functions in the context of holography, following \cite{Jottar:2010vp}\cite{Hyun:2011qj}.
In the rest of the introduction, we briefly explain aging and logarithmic conformal
field theories (LCFT).

Aging is one of the simplest nonequilibrium phenomena of many body systems, which is realized
when the system is rapidly brought out of equilibrium.
The basic properties of aging can be characterized by two-time correlation functions
with two-time scales, ``waiting time,'' when the system is perturbed after brought out of the
equilibrium, and ``response time,'' when the perturbation is measured.
The correlation functions show the typical behaviors: older systems with a bigger waiting time
relax in a slower manner than younger systems with smaller waiting time, see {\it e.g.}
\cite{HenkelBook2}\cite{HinrichsenReviewDP}\cite{Henkel:2007nept} for reviews.

In the context of Anti-de Sitter space / Conformal field theory correspondence (AdS/CFT) \cite{Aharony:1999ti},
the geometric realization of aging is put forward in \cite{Jottar:2010vp} 
by generalizing Schr\"odinger background \cite{Son:2008ye}\cite{Balasubramanian:2008dm}
with explicit time dependent terms.%
\footnote{See also \cite{Nakayama:2010xq} for some general time dependent deformations
of Schr\"odinger backgrounds.}  
The authors of \cite{Jottar:2010vp} observe that aging algebra can be realized
as a subset of Schr\"odinger algebra using a singular time dependent coordinate transformation.
Thus it is established that the time translation symmetry is broken globally, and the aging
symmetry is realized as conformal Schr\"odinger symmetry modulo time translation symmetry.
Yet, it is claimed that the time itself necessarily has to be complex to show a relaxation
process, which is one of the central properties of the aging \cite{Jottar:2010vp}.
This complexification of the time is certainly not preferred.

This drawback is overcome in \cite{Hyun:2011qj}, by introducing some decay modes of the bulk scalar field
along the `internal' spectator direction $\xi $, which is not explicitly visible from the dual
field theory with Sch\"odinger and aging symmetry. The resulting two-time correlation functions show a dissipative
behavior and exhibit the characteristic features of the aging system: power law decay,
broken time translation invariance and dynamical scaling between the time and spatial directions
\cite{Hyun:2011qj}, see also {\it e.g.} \cite{HenkelBook2}\cite{HinrichsenReviewDP}\cite{Henkel:2007nept}
from the field theory point of view. These properties are shown in the geometric realizations
of the aging background in the context of \cite{Son:2008ye}\cite{Balasubramanian:2008dm} as well as
of the aging in light-cone in the context of \cite{Goldberger:2008vg}\cite{Barbon:2008bg}.
The corresponding finite temperature generalizations for these two different backgrounds with
asymptotic aging invariance are also studied in \cite{Hyun:2011qj}.

LCFT is a CFT, in which some correlation functions have logarithmic singularities. For the review on LCFT, see, for example,  \cite{Flohr:2001zs}\cite{Gaberdiel:2001tr}.
Recently, there have been interesting developments in LCFT as a dual CFT of higher derivative gravity model.
It started in three dimensional gravity models \cite{Grumiller:2008qz}-\cite{Alishahiha:2010bw}
and then in their higher dimensional analogues, namely, critical gravity \cite{Lu:2011zk}-\cite{Lu:2011ks}.
Generically, these theories include  higher derivative terms as well as Einstein-Hilbert
and cosmological constant terms and admit the AdS spacetime as a vacuum solution.
In four and higher dimensional spacetime, the higher derivative terms are given
by some combination of $R^2$ and $R_{\mu\nu}R^{\mu\nu}$ terms.
When the coupling of the higher derivative terms is tuned,
the linearized perturbations of the metric around the AdS vacuum solution include
the massless and logarithmic modes \cite{Lu:2011zk}.%
\footnote{In the theory without these fine-tuned coupling of the higher derivative terms,
the fluctuations around AdS spacetime include the massive and massless graviton modes,
one of which is ghost-like. Since AdS spacetime has timelike boundary,
one may impose a boundary condition and truncate the ghost-like massive mode
if it falls off more slowly than the massless one\cite{Maldacena:2011mk}\cite{Lu:2011ks}.
It was shown in \cite{Hyun:2011ej}\cite{Hyun:2012mh} that the theory with the truncation becomes
effectively identical to the usual Einstein gravity at the classical but full nonlinear level
and thus the correlation functions of the dual CFT can be readily read off from those in the CFT dual
to Einstein gravity.  If we choose  special values on the coupling of higher derivative terms,
the massive and massless modes become degenerate and turn into the massless
and the logarithmic modes \cite{Lu:2011zk}.}
This is called critical gravity and its dual CFT becomes LCFT which includes stress-energy tensor
and its logarithmic counterpart.

The holographic descriptions of LCFT have been attempted
since early days of the AdS/CFT correspondence  using a pair of scalar fields \cite{Ghezelbash:1998rj}\cite{Kogan:1999bn}\cite{Myung:1999nd}.  Recently, the correlation functions of a pair of scalar operators in non-relativistic LCFT have  been considered  in the context of AdS/CFT correspondence using the Lifshitz background
\cite{Bergshoeff:2011xy}. In this paper, we study the correlation functions of a pair of scalar operators, which would play the role of order parameters, in the non-relativistic LCFT with Schr\"odinger and aging invariance.

We introduce a new type of action with
Schr\"odinger invariance and generalize the correlation function to obtain the logarithmic
corrections in detail in section \S \ref{sec:SchrLCFT}.
This is generalized for the system with aging invariance in section
\S \ref{sec:AgingLCFT}.
In section \S \ref{sec:Application}, we investigate various physical properties of the logarithmic
aging correlation functions and also seek possibilities to connect to the physical systems with
surface growth \cite{TakeuchiPRL}\cite{TakeuchiSciRep}
and their extension with logarithmic corrections \cite{Henkel:2011NP}.
We conclude in section \S \ref{sec:Conclusion}.

\section{Schr\"odinger LCFT}    \label{sec:SchrLCFT}

Let us consider some strongly coupled $ (2+1) $ dimensional dual field theory with Schr\"odinger symmetry,
living at the boundary of the Schr\"odinger background
	\begin{align}
	ds_u^2 =   \frac{L^2}{u^2} \left(  dy_1^2 + dy_2^2  -2 dx^{+}  dx^{-}
	- \frac{\gamma}{u^2} ~dx^{+2}  +  du^2  \right)  \;,
	\end{align}
where $\gamma$ indicates the fact that we consider two different cases on the same footing,
the Schr\"odinger background with $\gamma \neq 0$ \cite{Son:2008ye}\cite{Balasubramanian:2008dm}
and the AdS in light-cone $\gamma=0$ \cite{Goldberger:2008vg}\cite{Barbon:2008bg} at zero temperature.
It turns out that these two geometric realizations give the same physical properties for the
correlation functions at zero temperature for the scalar operators of given conformal dimensions.
Their finite temperature generalizations have been realized in
\cite{Herzog:2008wg}-\cite{Ammon:2010eq}
for $\gamma \neq 0$ and in \cite{Kim:2010tf}\cite{Kim:2010zq} for $\gamma=0$.
It is also observed that there exist clear distinctions in metric fluctuations between these two
backgrounds \cite{Hyun:2011qj}.

Motivated by the recent interests on LCFT from the holographic point of view \cite{Bergshoeff:2011xy},
we consider two scalar fields $\phi$ and $\tilde \phi$ in this background, whose action has the
following form
	\begin{align}  \label{SchrAction}
		S = K \int d^4 x \int_{u_B}^{\infty} du \sqrt{-g}
		& \left(  g^{MN} ~\partial_M \tilde \phi ~\partial_N \phi
		+ m^2 ~\tilde  \phi ~\phi +\frac{1}{2L^2}\tilde \phi^2 \right)  \;,
	\end{align}
where $u_B $ represents a cutoff near the boundary, $K = -\pi^3 L^5 /4 \kappa_{10}^2$,
$M, N = +, -, u, y_1, y_2$ and  $ \mu, \nu = +, y_1, y_2$.
Upon integrating out the field $\tilde \phi $, one can obtain an equivalent action with
second order time derivatives and fourth order spatial derivatives.
In the context of AdS/CFT, the logarithmic extension on Lifshitz backgrounds can be found
in \cite{Bergshoeff:2011xy}, where one can find some general comments for this action.

Let us comment the general solution space of the theory described by the action (\ref{SchrAction}).
In \cite{Bergshoeff:2011xy}, a bigger solution space for the Lifshitz theory
with two spin $0$ fields $\phi$ and $\tilde \phi $ with different parameters $m$ and $\tilde m$
is suggested. When specialized with $m = \tilde m$, the theory describes the Lifshitz LCFT.
It is straight forward to follow this features and generalize the solution space of the theory
by introducing two different parameters $m$ and $\tilde m$ for the fields $ \phi$ and $\tilde \phi$.

Contrasted to Lifshitz theories, Schr\"odinger theories have additional features
due to the isometry associated with the coordinate $x^- $, called ``dual particle mass.''
This can be shown explicitly as $\partial_-\phi=-{\cal M}\phi$ and
$\partial_-\tilde\phi=- \tilde {\mathcal M}\tilde\phi$ considering the fields $\phi$
and $\tilde \phi $ with two different values of $ \mathcal M$ and $\tilde {\mathcal M}$, respectively.%
\footnote{As mentioned in the introduction, we are concerned with the aging properties
which shows dissipative properties in the correlation functions. Thus we consider only the
decaying modes of the scalar fields $\phi $ and $\tilde \phi $. For the detailed discussion
for the differences between the decaying modes and oscillating modes, see  \cite{Hyun:2011qj}.}
Thus the solution space of our fields $\phi$ and $\tilde \phi$ can be described
by the parameters $(m, \mathcal M)$ and $(\tilde m, \tilde {\mathcal M})$.
In this paper, we consider the case $m = \tilde m$ and $\mathcal M = \tilde {\mathcal M}$,
which gives the Schr\"odinger LCFT.

The linearized field equations for $\phi$ and $\tilde \phi$ of the action (\ref{SchrAction}) become
	\begin{align}   \label{DiffEqSchr}
		&2{\cal M} \partial_+ \phi =
		- \frac{1}{u^2} \mathcal D  \phi - \vec \nabla ^2  \phi + \frac{1}{u^2} \tilde \phi \;, \\
		&2{\cal M} \partial_+ \tilde \phi =
		- \frac{1}{u^2} \mathcal D \tilde \phi - \vec \nabla ^2 \tilde \phi  \;,
	\end{align}
where
	\begin{align} \label{DiffOpSchr}
		\mathcal  D = u^2 \bigg[ \partial_u^2 - \frac{3}{u} \partial_u
		+ \frac{\gamma {\cal M}^2 - m^2 L^2 }{u^2} \bigg] \;,
	\end{align}
which is a differential operator for the Schr\"odinger case.

Following \cite{Hyun:2011qj}\cite{Bergshoeff:2011xy},
we construct the bulk to boundary Green's function $G_{ij} (u, \omega, \vec{k})$ as
	\begin{align}   \label{BulkToBoundaryGreensFunction}
		&\phi (u,x^\mu) = \int \frac{d^2 k}{(2\pi )^2}\frac{d \omega}{2 \pi} e^{i k_\mu x^\mu }
		\left[ G_{11} (u, k_\mu) J(k_\mu)+  G_{12} (u, k_\mu)\tilde J(k_\mu)\right]\;, \nonumber  \\
		&\tilde  \phi (u,x^\mu) = \int \frac{d^2 k'}{(2\pi )^2}\frac{d \omega'}{2 \pi} e^{-i k'_\mu x^\mu }
		\left[  G_{21} (u, k'_\mu) J(k'_\mu)+ G_{22} (u, k'_\mu) \tilde J(k'_\mu)\right]\;,
	\end{align}
where we use $x^\mu = (x^+, \vec y), ~k_\mu = (-\omega, \vec k) $, which give
$ k_\mu x^\mu = \vec{k} \cdot \vec{y} - \omega x^+ $.
We choose $G_{21}=0$, which is in accord with the structure of the equations of motion given
in (\ref{DiffEqSchr}). The Green's functions satisfy
	\begin{align}   \label{3DifferentialEquations}
	\mathcal D G_{11} &= 0  \;, \qquad
	\mathcal D G_{12} = G_{22}   \;, \qquad
	\mathcal D G_{22} = 0   \;,
	\end{align}
where $\mathcal D$ is defined in (\ref{DiffOpSchr}) and $q = \sqrt{\vec k^2 +2 {\cal M} i\omega}$.
The Green's functions $G_{11}$ and $G_{22}$ are given by
	\begin{align}		
		&G_{11}(u, k_\mu)=  c_{11}~u^{2} K_\nu (q u)\;, \qquad
		G_{22}(u, k_\mu)=  c_{22}~u^{2} K_\nu (q u) \;,
	\end{align}
where $\nu = \sqrt{4+ m^2L^2 -\gamma {\mathcal M}^2} $.
The normalization constants $c_{11}=c_{22}= \frac{1}{u_B^{2} K_\nu (q u_B)}$ can be determined
by requiring that $G_{11}(u_B, k_\mu) =G_{22}(u_B, k_\mu) =1$ \cite{Son:2002sd}.

There exists another Green's function $G_{12}$ due to the action we choose (\ref{SchrAction}),
which satisfies
	\begin{align}
	\mathcal D G_{12} = G_{22}   \;.
	\end{align}
To evaluate $G_{12}$, we use the same methods used in \cite{Kogan:1999bn}.
Using
    \begin{eqnarray}
        \left[ \mathcal D \,, \frac{d}{d \nu} \right]  = 2 \nu \;,
    \end{eqnarray}
and the fact that $ \mathcal D G_{22}=0 $, we get
    \begin{eqnarray}
        \mathcal D \left( \frac{1}{2 \nu} \frac{d}{d\nu} G_{22} \right) =G_{22} \;.
    \end{eqnarray}
Thus
	\begin{align}   \label{G12Expression}
        G_{12} = \frac{1}{2\nu} \frac{d}{d \nu} G_{22}
        = \frac{1}{2\nu} \frac{d}{d \nu}\left( \frac{u^{2} K_{\nu} (q u)}
        {u_{B}^{2} K_{\nu} (q u_{B})} \right) \;.
    \end{align}
We will not write down the explicit expression, as it is lengthy and not so illuminating.

After plugging the bulk equation of motion into the action (\ref{SchrAction}),
the boundary action becomes of the form
	\begin{align}   \label{boundaryAction}
		S_B &= -K \int  d^3 x \frac{L^5}{u^{5}} ~\tilde \phi  \frac{u^2}{L^2} \partial_u
			 \phi \Big|_{u_B} \nonumber \\
			&= -K \int d x^+ \theta (x^+) \int \frac{d \omega}{2 \pi} \frac{d \omega'}{2 \pi}
			e^{-i (\omega - \omega') x^+} \int d^2 y  \int \frac{d^2 k}{(2\pi )^2}  \frac{d^2 k'}{(2\pi )^2}
			e^{i (\vec{k} - \vec{k'}) \cdot \vec{y}}   \nonumber \\
			&\qquad \times \tilde J (k'_\mu) \Big(\mathcal F_1 (u_B, k'_\mu, k_\mu ) J (k_\mu)
			+ \mathcal F_2 (u_B, k'_\mu, k_\mu) \tilde J (k_\mu) \Big) \Big|_{u_B} \;.
	\end{align}
For the Schr\"odinger case, the system has local time translation invariance, thus the time integral is
trivially evaluated to give delta function.
The ${\cal F}$'s are given by
	\begin{align}    \label{F1F2}
		{\cal F}_1	&= \frac{L^3}{u^3}G_{22} (k'_\mu)~ \partial_u ~ G_{11} (k_\mu)  \;, \\
		{\cal F}_2	&= \frac{L^3}{u^3}G_{22} (k'_\mu) ~\partial_u ~ G_{12} (k_\mu)   \;.
		\label{F1F2B}
	\end{align}

${\mathcal F}_1$ can be expressed, using
	\begin{align}
		\partial_u [u^2 K_\nu (q u)] = u [(2-\nu) K_\nu (q u) -qu K_{\nu-1} (qu)] \;, 	
	\end{align}	
as
	\begin{align}   \label{F1functionSchr}
	 	{\cal F}_1 	= \frac{L^3}{u_B^4} \left[(2-\nu)-q u_B \frac{K_{\nu-1} ( q u_B) }{K_{\nu} ( q u_B) }  \right] 
		\approx  -\frac{2 \Gamma(1-\nu)}{\Gamma(\nu)} \frac{L^3}{u_B^4}
		\left( \frac{ q u_{B} }{2} \right)^{2 \nu}  \;.
	\end{align}
Note that we keep only the first non-trivial contribution in the small $u_B $ expansion and
use the normalization for the wave function to be normalized to unity at the boundary \cite{Son:2002sd}.
This result is already noticed in several different places including
\cite{Balasubramanian:2008dm}\cite{Goldberger:2008vg}\cite{Hyun:2011qj}.
The momentum space correlation function is given by
\begin{align}
	&\langle \tilde {\phi} (\omega',\vec{k}') {\phi}(\omega, \vec{k}) \rangle
	= -2 (2\pi)^{-3}  \delta (\vec{k'} - \vec{k} )~\delta (\omega' -\omega)
	~ {\cal F}_1 (u_{B},\omega,\vec{k}) \;,
\end{align}
where we use $ {\phi}(x^+, \vec{y}) $ as the dual operator of the bulk field
$ {\phi}(u, x^+, \vec{y}) $.

The coordinate space correlation function can be computed using the inverse Fourier transformation
	\begin{align}   \label{ZeroTCorrelatorSchr}
		&\langle \tilde {\phi} (x_{2}^{+},\vec{y}_{2})  {\phi}(x_{1}^{+},\vec{y}_{1}) \rangle
		= \int \frac{d \omega'}{2 \pi} \frac{d^2 k'}{(2\pi )^2} \frac{d \omega}{2 \pi} \frac{d^2 k}{(2\pi )^2}
		e^{-i \vec{k'} \cdot \vec{y}_{2}+i \vec k \cdot \vec{y}_{1}}
		e^{ i \omega' \cdot x^{+}_{2} - i \omega \cdot x_1^+}
		\langle \tilde \phi (\omega',\vec{k}') \phi(\omega, \vec{k}) \rangle  \nonumber \\
		&\qquad =  \frac{\Gamma (1-\nu)}{ \Gamma (\nu) \Gamma (-\nu)}
		\frac{L^3 \mathcal M^{1+\nu}  }{\pi 2^{\nu-1}  u_B^{4-2\nu}}
		\cdot \frac{\theta (x_2^+) \theta(x_2^+ - x_1^+) }{ (x_2^+ - x_1^+)^{2+\nu}}
		\cdot \exp \left( -{ \frac{\mathcal M (\vec y_2 - \vec y_1)^2 }{2 ( x_2^+ - x_1^+)}}\right)  \;.
	\end{align}
This is the result we are looking for.
Note that the final result has the same form as that of the Schr\"odinger case which is given in \cite{Hyun:2011qj}.

Now let us turn to evaluate the ${\cal F}_2$ using (\ref{G12Expression})
	\begin{align}  \label{ZeroTCorrelatorSchrLog}
		&\langle \phi (x_{2}^{+},\vec{y}_{2}) \phi (x_{1}^{+},\vec{y}_{1}) \rangle
		= 	-2 \theta (x_2^+)~ \int\frac{d\omega}{2\pi} \frac{d^2 k}{(2\pi )^2}
		e^{-i \vec{k} \cdot (\vec y_2-\vec y_1)} e^{ i \omega (x^+_2-x_1^+) }
		{\cal F}_2 (u_{B},\omega,\vec{k})  \nonumber \\ 
		&\quad = \frac{1}{ \Gamma (\nu)}
		\frac{L^3 \mathcal M^{1+\nu}  }{\pi 2^{\nu-1}  u_B^{4-2\nu}}
		\cdot \frac{\theta (x_2^+) \theta(x_2^+ - x_1^+) }{ (x_2^+ - x_1^+)^{2+\nu}}
		\cdot \exp \left( -{ \frac{\mathcal M (\vec y_2 - \vec y_1)^2 }{2 ( x_2^+ - x_1^+)}}\right) \nonumber \\
		&\qquad \times \left( 1 -\frac{\Gamma (\nu)'}{\Gamma (\nu)} + \nu \ln [\frac{M u_B^2}{2(x_2^+ - x_1^+)}]
		 \right) \;.
	\end{align}
This is our main result in this section.
The correlation function of the dual field theory acquires the logarithmic time dependence,
which is very interesting. We give comments on it after we get more general results in the aging case below.

\section{Aging LCFT}    \label{sec:AgingLCFT}

In the previous section, we evaluate the two point correlation function of scalar order parameters
of the dual field theory with Schr\"odinger symmetry. Similarly, we would like to compute
the correlation functions with aging symmetry, which is intrinsically out of equilibrium and time dependent.

Let us consider the following aging background
	\begin{align}
		ds_u^2 =   \frac{L^2}{u^2} \left(  dy_1^2 + dy_2^2  -2 dx^{+}  dx^{-}
		- \left(\frac{\gamma}{u^2} +  \frac{\alpha}{x^+} \right) dx^{+ 2}
			+ \frac{2 \alpha  }{u} du dx^+  +  du^2  \right) \;,
	\end{align}
whose construction is explained in	\cite{Jottar:2010vp}\cite{Hyun:2011qj}.
The parameter $\alpha $ signifies the fact that the system has an explicit time dependence, while
the parameter $\gamma $ represents the difference between the aging in Schr\"odinger background
with $\gamma = 1$ and the aging in light-cone with $\gamma = 0$.
The system is one of the simplest time dependent background, with which we can explicitly explore
important physical properties related to out of equilibrium. The reader can find more detailed
information in \cite{Jottar:2010vp}\cite{Hyun:2011qj}.

Consider two scalar fields $\phi$ and $\tilde \phi$ whose action is given by (\ref{SchrAction}).
The linearized field equations for $\phi$ and $\tilde \phi$ become
	\begin{align}
		&2{\cal M} \left( \partial_+ + \frac{\alpha  {\cal M}}{2 x^+}   \right) \phi =
		- \frac{1}{u^2} \mathcal D_\alpha  \phi - \vec \nabla ^2  \phi + \frac{1}{u^2} \tilde \phi \;, \\
		&2{\cal M} \left( \partial_+ + \frac{\alpha  {\cal M}}{2 x^+}   \right) \tilde \phi =
		- \frac{1}{u^2} \mathcal D_\alpha \tilde \phi - \vec \nabla ^2 \tilde \phi  \;,
	\label{fieldEQphi}
	\end{align}
where we treat $x^-$ coordinate special by assigning $\partial_-\phi=-{\cal M}\phi$ and
$\partial_-\tilde\phi=- {\cal M}\tilde\phi$ and use
	\begin{align}   \label{DiffOperator2}
		\mathcal  D_\alpha = u^2 \bigg[ \partial_u^2 - \frac{2 {\cal M}\alpha +3}{u} \partial_u
		+ \frac{4{\mathcal M}\alpha + \alpha^2 {\mathcal M}^2 + \gamma {\mathcal M}^2 - m^2 L^2 }{u^2} \bigg] \;,
	\end{align}
which is a modified differential operator for the aging case.

The bulk to boundary Green's function $G^\alpha_{ij} (u, \omega, \vec{k})$ may be defined similar to the
Schr\"odinger case given in (\ref{BulkToBoundaryGreensFunction}).
	\begin{align}   \label{BulkToBoundaryGreensFunctionAging}
		&\phi (u,x^\mu) = \int \frac{d^2 k}{(2\pi )^2}\frac{d \omega}{2 \pi} e^{i k_\mu x^\mu } c
		\left[ G^\alpha_{11} (u, k_\mu) J(k_\mu)+  G^\alpha_{12} (u, k_\mu)\tilde J(k_\mu)\right]\;,   \\
		&\tilde  \phi (u,x^\mu) = \int \frac{d^2 k'}{(2\pi )^2}\frac{d \omega'}{2 \pi} e^{-i k'_\mu x^\mu } c
		\left[  G^\alpha_{21} (u, k'_\mu) J(k'_\mu)+ G^\alpha_{22} (u, k'_\mu) \tilde J(k'_\mu)\right]\;.
	\end{align}
Once again, we choose $G^\alpha_{21}=0$ and the Green's functions satisfy
	\begin{align}
	\mathcal D_\alpha G^\alpha_{11} &= 0  \;, \qquad
	\mathcal D_\alpha G^\alpha_{12} = G^\alpha_{22}   \;, \qquad
	\mathcal D_\alpha G^\alpha_{22} = 0   \;,
	\end{align}
where $\mathcal D_\alpha$ is defined in (\ref{DiffOperator2}).
We find that $G^\alpha_{ij} = \left(\frac{\alpha x^+}{u^{2}} \right)^{-\alpha \mathcal M} G_{ij} $,
where $ G_{ij}$ satisfies equations (\ref{3DifferentialEquations}) - (\ref{G12Expression})
of the Schr\"odinger case.
	
After plugging the bulk equation of motion, the boundary action becomes of the form
	\begin{align}  \label{AgingBoundaryAction}
		S_B &= -K \int  d^3 x \frac{L^5}{u^{5}} ~\tilde \phi \left( \frac{u^2}{L^2} \partial_u
			- \frac{\alpha \mathcal M u }{L^2} \right) \phi \Big|_{u_B} \nonumber \\
			&= -K \int d x^+ \theta (x^+) \int \frac{d \omega}{2 \pi} \frac{d \omega'}{2 \pi}
			e^{-i (\omega - \omega') x^+} \left(\frac{\alpha x^+}{u^{2}} \right)^{-\alpha \mathcal M}
			\int d^2 y  \int \frac{d^2 k}{(2\pi )^2}  \frac{d^2 k'}{(2\pi )^2}
			e^{i (\vec{k} - \vec{k'}) \cdot \vec{y}}  \nonumber \\
			&\qquad \times  \tilde J (k'_\mu)
			\Big(\mathcal F_1 (u_B, k'_\mu, k_\mu ) J (k_\mu)
			+ \mathcal F_2 (u_B, k'_\mu, k_\mu) \tilde J (k_\mu) \Big) \Big|_{u_B} \;,
	\end{align}
where ${\cal F}$'s are given by the equations (\ref{F1F2}) and (\ref{F1F2B}).

Note that $\theta (x^+) $ represents our restriction on the region of time to be $x^+ \geq 0$, which is
very natural for our setup from the metric point of view.	The physical significances and effects of
the time boundary and associated boundary condition are explained in detail in \cite{Hyun:2011qj}.

To evaluate the momentum space correlation functions, we first evaluate the time integral
for the range $0 \leq x^+ < \infty$ and define
$ {\cal G} (w) =  \int dx^+ e^{i w x^+} \theta (x^+) \cdot  (\alpha x^+)^{-\alpha \mathcal M}$. Then
	\begin{align}
		{\cal G} (w) =\alpha^{-\alpha M_I} |w|^{-1+\alpha \mathcal M} \Gamma (1-\alpha \mathcal M)
		\left(- i \cos (\frac{\pi \alpha \mathcal M}{2}) \text{sign} (w)
		+ \sin (\frac{\pi \alpha \mathcal M}{2})\right) \;.
	\label{GMI}
	\end{align}
where $w=\omega' - \omega$.
These results are the same as those evaluated by Laplace transform.

Thus the momentum space correlation function is given by
\begin{align}
	&\langle \tilde {\phi} (\omega',\vec{k}') {\phi}(\omega, \vec{k}) \rangle
	= -2 (2\pi)^{-3}  \delta (\vec{k'} - \vec{k} )~ {\cal G} (\omega' -\omega)
	~ {\cal F}_1 (u_{B},\omega,\vec{k}) \;,
\end{align}
where ${\cal F}_1$ is given by (\ref{F1functionSchr}).
We use the spatial translational invariance to get the delta function for $\vec k $,
but not for $\omega$ due to the lack of the time translation symmetry.

Let us calculate the coordinate space correlation function as
	\begin{align}
		&\langle \tilde {\phi} (x_{2}^{+},\vec{y}_{2})  {\phi}(x_{1}^{+},\vec{y}_{1}) \rangle \\
		&\quad = \int \frac{d \omega'}{2 \pi} \frac{d^2 k'}{(2\pi )^2} \frac{d \omega}{2 \pi} \frac{d^2 k}{(2\pi )^2}
		e^{-i \vec{k'} \cdot \vec{y}_{2}+i \vec k \cdot \vec{y}_{1}}
		e^{ i \omega' \cdot x^{+}_{2} - i \omega \cdot x_1^+}
	\left( \alpha^2 x_1^+ x_2^+ \right)^{\frac{\alpha \cal M}{2}}
	\langle \tilde \phi (\omega',\vec{k}') \phi (\omega, \vec{k}) \rangle \;. \nonumber
	\end{align}
After carrying out the integral for the $\omega' $ first
	\begin{align}
  		\int \frac{d \omega'}{2 \pi} ~e^{  i ( \omega' - \omega)  x^{+}_{2}} ~ {\cal G} (\omega' -\omega)
 		= \theta (x_2^+)~\alpha^{-\alpha \cal M}  |x_2^+|^{- \alpha \cal M} \;,
	\end{align}
the expression reduces to the Schr\"odinger case in the time independent part, whose calculation is done similarly.
Finally we obtain
	\begin{align}    	\label{ZeroTCorrelatorAging}
		&\langle \tilde{\phi} (x_{2}^{+},\vec{y}_{2}) {\phi}(x_{1}^{+},\vec{y}_{1}) \rangle
		= 	-2 \theta (x_2^+)~\left( \frac{x_2^+}{x_1^+} \right)^{-\frac{\alpha \mathcal M}{2} }
		\int\frac{d\omega}{2\pi} \frac{d^2 k}{(2\pi )^2}
		e^{-i \vec{k} \cdot (\vec y_2-\vec y_1)} e^{ i \omega (x^+_2-x_1^+) }
		{\cal F}_1 (u_{B},\omega,\vec{k})  \nonumber \\
		&\qquad = \frac{-\nu}{ \Gamma (\nu)} \frac{L^3 \mathcal M^{1+\nu}  }{\pi 2^{\nu-1}  u_B^{4-2\nu}}
		\cdot \frac{\theta (x_2^+) \theta(x_2^+ - x_1^+) }{ (x_2^+ - x_1^+)^{2+\nu}}
		\cdot \left( \frac{x_2^+}{x_1^+} \right)^{-\frac{\alpha \mathcal M}{2} }   \cdot \exp \left( -{ \frac{\mathcal M (\vec y_2 - \vec y_1)^2 }{2 ( x_2^+ - x_1^+)}}\right)  \;.
	\end{align}
This result has exactly the same form as that of the aging case computed in \cite{Hyun:2011qj}\cite{Jottar:2010vp}.%
\footnote{
We have a slightly different expression compared to \cite{Hyun:2011qj} using the identity of
the $\Gamma $ function. This form is slightly more convenient for the evaluation of $\mathcal F_2 $.
}

We can similarly evaluate $ \langle \phi \phi \rangle $ correlation function as
\begin{align}  \label{ZeroTCorrelatorAgingLog}
		&\langle {\phi} (x_{2}^{+},\vec{y}_{2})  {\phi}(x_{1}^{+},\vec{y}_{1}) \rangle \nonumber  \\
		&= \frac{1}{ \Gamma (\nu) }
		\frac{L^3 \mathcal M^{1+\nu}  }{\pi 2^{\nu-1}  u_B^{4-2\nu}}
		\cdot \frac{\theta (x_2^+) \theta(x_2^+ - x_1^+) }{ (x_2^+ - x_1^+)^{2+\nu}}
		\cdot \left( \frac{x_2^+}{x_1^+} \right)^{-\frac{\alpha \mathcal M}{2} }   \cdot
		\exp \left( -{ \frac{\mathcal M (\vec y_2 - \vec y_1)^2 }{2 ( x_2^+ - x_1^+)}}\right) \nonumber \\
		&\qquad \times \left( 1 -\frac{\Gamma (\nu)'}{\Gamma (\nu)}
		+ \nu \ln \left[\frac{M u_B^2}{2(x_2^+ - x_1^+)}\right]  \right) \;.
	\end{align}
This expression is our main result, which has all the essential properties related to the
logarithmic extension of the aging holography.
Let us comment about this coordinate space correlation function. 	
	\begin{itemize}
		\item There are clear differences between the two functions
		$\langle \tilde{\phi} (x_{2}^{+},\vec{y}_{2})  {\phi}(x_{1}^{+},\vec{y}_{1}) \rangle $
		and $\langle {\phi} (x_{2}^{+},\vec{y}_{2}) $  $ {\phi}(x_{1}^{+},\vec{y}_{1}) \rangle $.
		The latter can be obtained simply by taking a derivative of the former with respect to
		the parameter $\nu $. This is clear from the structure of the differential operator.
		The same result can be obtained by a direct, yet lengthy, calculation outlined in the appendix
		\S \ref{sec:LogEvaluation}.
		
		\item The correlation function
		$\langle {\phi} (x_{2}^{+},\vec{y}_{2}) {\phi}(x_{1}^{+},\vec{y}_{1}) \rangle $ reveals an
		additional logarithmic dependence only of the time difference $\ln (x_2^+ - x_1^+) $, besides
		some irrelevant constant factors. This is the main feature of the aging LCFT from our bulk construction.
		Interestingly, the spatial dependence $ \vec y_2 - \vec y_1$ only shows up in the exponential part,
		which is responsible for the well defined decaying properties of the Schr\"odinger and aging
		correlation functions for both the large distance $\vec y_2 - \vec y_1 \rightarrow \infty  $
		and small time $x_2^+ - x_1^+ \rightarrow 0 $ regimes.
		
		\item We confirm the relation between the aging correlation functions and those of the Schr\"odinger
			\begin{align}
				&\langle \tilde {\phi} (x_{2}^{+},\vec{y}_{2})  {\phi}(x_{1}^{+},\vec{y}_{1}) \rangle_{Aging}
				= \left( \frac{x_2^+}{x_1^+} \right)^{-\frac{\alpha \mathcal M}{2} }
				\langle \tilde {\phi} (x_{2}^{+},\vec{y}_{2})  {\phi}(x_{1}^{+},\vec{y}_{1}) \rangle_{Schr} \;, \\
				&\langle {\phi} (x_{2}^{+},\vec{y}_{2})  {\phi}(x_{1}^{+},\vec{y}_{1}) \rangle_{Aging}
				= \left( \frac{x_2^+}{x_1^+} \right)^{-\frac{\alpha \mathcal M}{2} }
				\langle {\phi} (x_{2}^{+},\vec{y}_{2})  {\phi}(x_{1}^{+},\vec{y}_{1}) \rangle_{Schr} \;,			
			\end{align}
		which is one of the important properties of aging observed in \cite{Jottar:2010vp}\cite{Hyun:2011qj}.
		
		\item We stress that the final form of the correlation function
        with logarithmic extension is fixed by the backgrounds motivated by string theory solutions. 
        Schr\"odinger and aging backgrounds can be  constructed
        directly from the AdS$_5 \times S^5 $ solution using the null Melvin twist
        \cite{Gimon:2003xk}\cite{Alishahiha:2003ru} for $\gamma \neq 0$
        and the light-cone reduction \cite{Maldacena:2008wh}\cite{Kim:2010tf} for $\gamma=0$.
        Attempts to construct the corresponding dual field theories have been put forward, {\it e.g.} in
        \cite{Adams:2008wt}\cite{Hyun:2011qj}. 
	\end{itemize}

\section{Physics of two-time correlation and response functions}    \label{sec:Application}

We obtained the correlation functions and their logarithmic extensions of the Schr\"odinger case in
\S \ref{sec:SchrLCFT} and aging case in \S \ref{sec:AgingLCFT}. In this section, we describe the
physical properties of them in detail and seek their applications to the KPZ universality class,
motivated by its recent theoretical extension to the aging regime \cite{Henkel:2011NP}.
Aging cases offer more general class of correlation functions with explicit breaking of time
translational invariance compared to the Schr\"odinger case. Thus we concentrate on the aging phenomena and
comment on the Schr\"odinger case briefly.

Let us first comment about the physical identifications of $ \langle \tilde {\phi}  {\phi} \rangle $
and $ \langle {\phi}  {\phi} \rangle $. From the point of view of our detailed construction
outlined in \S \ref{sec:SchrLCFT}, these are naturally identified as correlation functions
of the logarithmic pair  $ \phi$ and $ \tilde \phi $.
Yet, it is compelling to identify them as response functions by coupling the logarithmic pair
$ (\phi, \tilde \phi) $ with some external fields $ (h, \tilde h) $ as
$\delta \mathcal L = h \cdot \phi + \tilde h \cdot \tilde \phi $. Starting from one point function
$ \langle {\phi} \rangle $, the variations with respect to $ \tilde h $ and $h $ will give the response
functions $ \langle \tilde {\phi}  {\phi} \rangle $ and $ \langle {\phi}  {\phi} \rangle $, respectively.
Similar identification in the field theory side is already adapted in the context of aging
and its logarithmic extension \cite{Henkel:2011NP}\cite{Henkel:2010hd}.
Thus we would like to identify $ \langle \tilde {\phi}  {\phi} \rangle $ and
$ \langle {\phi}  {\phi} \rangle $ as our response functions.
In this section, we freely take the view them as correlation functions or response functions.
We investigate the properties of $ \langle \tilde {\phi}  {\phi} \rangle $ and
$ \langle {\phi}  {\phi} \rangle $ in this section and try to seek possible connections to KPZ class.

\subsection{Correlation function }

The two-time correlation function of the geometric realizations with aging isometry
is identified and analyzed in the aging regime \cite{Hyun:2011qj}, reproducing the characteristic
features of its time dependence: slow power-law decay,
broken time translation invariance and dynamical scaling \cite{Henkel:2007nept}.
Here we would like to analyze the full profile of the correlation function {\it without} the
logarithmic extensions, motivated by
the recent interest in the field theory side \cite{Henkel:2011NP}\cite{Henkel:2010hd}.
As already noted above, the response function $ \langle \tilde {\phi}  {\phi} \rangle $ has the
same form as the correlation function obtained in \cite{Hyun:2011qj}. Thus we analyze its physical
properties here together.

Let us consider the two-time correlation function
$ C(x_2^+ , x_1^+) = \langle \tilde{\phi} (x_{2}^{+},\vec{y}_{1}) {\phi}(x_{1}^{+},\vec{y}_{1}) \rangle  $
obtained by identifying $ \vec{y}_{2} = \vec{y}_{1}$ from (\ref{ZeroTCorrelatorAging})
	\begin{align}   \label{TwoTimeCorrA}
		C(x_2^+ , x_1^+) &= C_0 (x_2^+ - x_1^+)^{-2-\nu}
		\cdot \left( \frac{x_2^+}{x_1^+} \right)^{-\frac{\alpha \mathcal M}{2} }  \;,
	\end{align}
where $ C_0 = \frac{-\nu}{ \Gamma (\nu)} \frac{L^3 \mathcal M^{1+\nu}  }{\pi 2^{\nu-1}  u_B^{4-2\nu}} $.
This depends on two different time scales, $ x_1^+ $ and
$  x_2^+ $. $ x_1^+ $ is called a {\it waiting time},
which is the time passed after the system is quenched or prepared.
Now the system is disturbed with small perturbations at this waiting time $ x_1^+ $,
and subsequently measured at $x_2^+ $ which is called a {\it response time}.
Explicit dependence on the waiting and response times in the two-time correlation function
is a crucial difference of the aging system.
The correlation functions of the Schr\"odinger invariant system or of the systems
with time translational invariance, in general, only depends on $x_2^+ - x_1^+ $,
the time difference between the response time and the waiting time.
These time dependent features manifest themselves in the term with $\alpha \mathcal M $ dependence.
The correlation function without the logarithmic corrections is explicitly
depicted in \cite{Hyun:2011qj} for the aging regime, which is $x_2^+ / x_1^+ \gg 1$
and $x_2^+, x_1^+ \gg 1 $. There, it is demonstrated that the plots show the slow power law dynamics
with explicit waiting time dependence,
in addition to the excellent collapse in the aging regime revealing the universal features.

Here we would like to explore the entire range of the waiting and response times
motivated by recent interesting applications \cite{Henkel:2011NP}\cite{Henkel:2010hd}.
It is more transparent to adopt the parametrization, $y = x_2^+ / x_1^+ $,  under which
	\begin{align}   \label{TwoTimeCorr}
		C(s, y) &= C_0~ s^{-2 -\nu} y^{-2-\nu-\frac{\alpha \mathcal M}{2}}
		\left(1- \frac{1}{y} \right)^{-2 -\nu} \;,
	\end{align}
where $s=x_1^+ $ is the waiting time, $1 \leq y \leq \infty$
and $\nu = \sqrt{4+m^2L^2 -\gamma \mathcal M^2}$.
Note that it is only a function of the scaling variable $y $ apart from the dependence of
the waiting time $s^{-2 -\nu} $.

\subsection{Correlation function, roughness and exponents}

Growth phenomena have attracted much attention due to their wide applications in physics and beyond.
Their examples include paper wetting, epidemics, flux lines in a superconductor and
fluid flow in porous media, see {\it e.g.}
\cite{BSBook}\cite{HinrichsenReviewDP}\cite{HenkelBook1}.
We would like to point out two characteristic properties of these phenomena,
which are relevant for the following discussions.
First, this is intrinsically nonequilibrium process and breaks the detailed balance \cite{HenkelBook1},
while the time independent treatments still provide much insight \cite{BSBook}.
Simply the system is out of equilibrium due to the fact that noise or randomness is an essential part of 
its dynamics. 
Second, the concept of the local scale invariance and associated universal exponents are useful descriptions,
which are similar to the equilibrium critical phenomena. The second property is particularly important
for our purpose. We consider the aging as well as the growth properties both for the equilibrium
and for the out of equilibrium systems, using the Schr\"odinger and aging holography, respectively.

Recently, Kardar-Parisi-Zhang type universality class in 1+1 dimensions is experimentally realized in the growing
interfaces of the turbulent liquid crystals \cite{TakeuchiPRL}\cite{TakeuchiSciRep}.
The roughness of interfaces is quantified by their width $w (l, t) $ defined as the standard deviation
of the interface height $h (x, t) $ over a length scale $l $ at time $t $, or equivalently the
radius $R (x, t) $ along the circle due to the fact that the universality class is independent of
the shapes of the interface \cite{TakeuchiSciRep}. The authors measure the interfacial width
$w (l, t) \equiv  \langle \sqrt{\langle [ h(x,t) - \langle h\rangle_l ]^2 \rangle_l } \rangle $
and the height-difference correlation function
$ C(l, t) \equiv \langle [ h(x+l,t) - h(x,t) ]^2 \rangle $
where $ \langle \cdots \rangle_l $ and $ \langle \cdots \rangle $ denote the average over a segment
of length $ l $ and all over the interface and ensembles, respectively \cite{TakeuchiPRL}.
Both $w (l, t)$ and $ C(l, t)^{1/2} $ are common quantities for characterizing the roughness,% 
\footnote{We concentrate on the correlation function due to the fact that both the $w (l, t)$ and $ C(l, t)^{1/2} $
show similar scaling properties, and the latter is readily available for the holographic theories.}
for which the so-called Family-Vicsek scaling
\cite{FVScaling} is expected to hold. The dynamical scaling property is
	\begin{align}
		 C(l, t)^{1/2} \sim t^{\mathrm b} F(l t^{-1/z}) \sim \left\{
		 \begin{array}{ccc}
			l^{\mathrm a} & \text{for} & l \ll l_* \\
			t^{\mathrm b} & \text{for} & l \gg l_*  \end{array}
		 \right. \;,
	\end{align}
with two characteristic exponents: the roughness exponent ${\mathrm a} $ and the growth exponent
${\mathrm b} $. The dynamical exponent is given by
$z= {\mathrm a} /{\mathrm b} $, and the cross over length scale is $ l_* \sim t^{1/z} $.
For an infinite system, the correlation function behaves as $C(l, t)^{1/2} \sim t^{\mathrm b} $
at large times.

In the holographic approach, the dynamical exponent $ z $ and the number of spatial directions $d $
are built-in. In our case, $ z=2 $ and $ d=2 $. Thus we can fix the roughness exponent ${\mathrm a} $ by
determining the growth exponent ${\mathrm b} $ from the limiting behavior of the correlation function,
$C(l, t)^{1/2} \sim t^{\mathrm b} $, at large times for our infinite system.
Once we take $ |\vec y_2 - \vec y_1 | = l  \ll t = x_2^+ - x_1^+ $, the correlation functions
(\ref{ZeroTCorrelatorSchr}) and (\ref{ZeroTCorrelatorAging}) behave as
	\begin{align}
		&\langle \tilde {\phi} (x_{1}^{+}+t,\vec l)  {\phi}(x_{1}^{+},0) \rangle_{Schr}
		\sim  t^{-2-\nu}  \;, \\
		&\langle \tilde {\phi} (x_{1}^{+}+t,\vec l)  {\phi}(x_{1}^{+}, 0) \rangle_{Aging}
		\sim  t^{-2-\nu -\frac{\alpha \mathcal M}{2}}  \;,		
	\end{align}
where we assume $ t \gg x_1^+ $ in the latter. Thus
	\begin{gather}
		{\mathrm b}_{Schr} = -1 - \nu / 2 \;,   \qquad {\mathrm a}_{Schr} = -2 - \nu \;,  \\
		{\mathrm b}_{Aging} = -1 - \nu / 2 - \alpha \mathcal M / 4 \;,
		\qquad {\mathrm a}_{Aging} =  -2 - \nu - \alpha \mathcal M / 2 \;,		
	\end{gather}
where the ratio between $ {\mathrm a}$ and ${\mathrm b} $ is fixed, $ {\mathrm a} / {\mathrm b}=z=2 $.%
\footnote{Surely, these exponents are not belong to the KPZ universality class 
(${\mathrm a} = 1/2 $ and ${\mathrm b} =1/3 $ in $1$ + $1$ dimensions), which is realized in
growing interfaces of the turbulent liquid crystals \cite{TakeuchiPRL}\cite{TakeuchiSciRep}.
It is necessary to study the holographic theory with $z=3/2 $ to directly contact with the KPZ class.
Our non-relativistic conformal case with $z=2 $ is more close to the Edwards-Wilkinson class\cite{EWclass}
with the same dynamical exponent.
}
These relations show that the growth exponent ${\mathrm b}_{Schr} $ in the Schr\"odinger case
is given by the parameter $\nu $, which measures the conformal weight of the dual scalar operator,
while the dynamical exponent is fixed from the construction as $z=2 $.
Similar story holds for the aging case, where the growth exponent ${\mathrm b}_{Aging} $
is a function of $\alpha \mathcal M $ in the metric, as well as $\nu $.
Note that there are interesting qualitative changes for $ \mathrm a = 0$ or $ \mathrm b = 0$,
which become clear below.

\begin{figure}[!ht]
\begin{center}
\begin{tabular}{cc}
	 \includegraphics[width=0.5\textwidth]{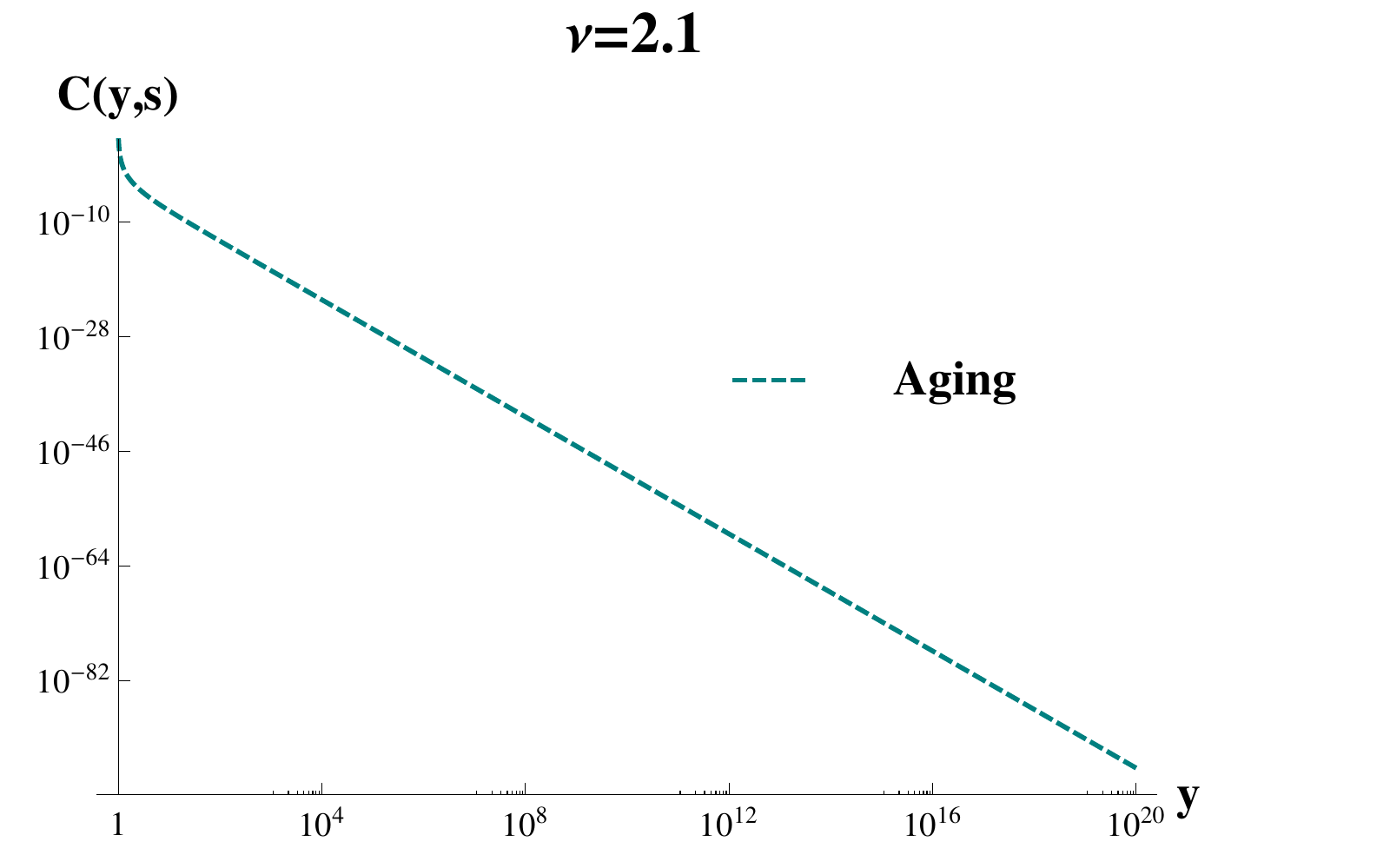} \quad
	 \includegraphics[width=0.5\textwidth]{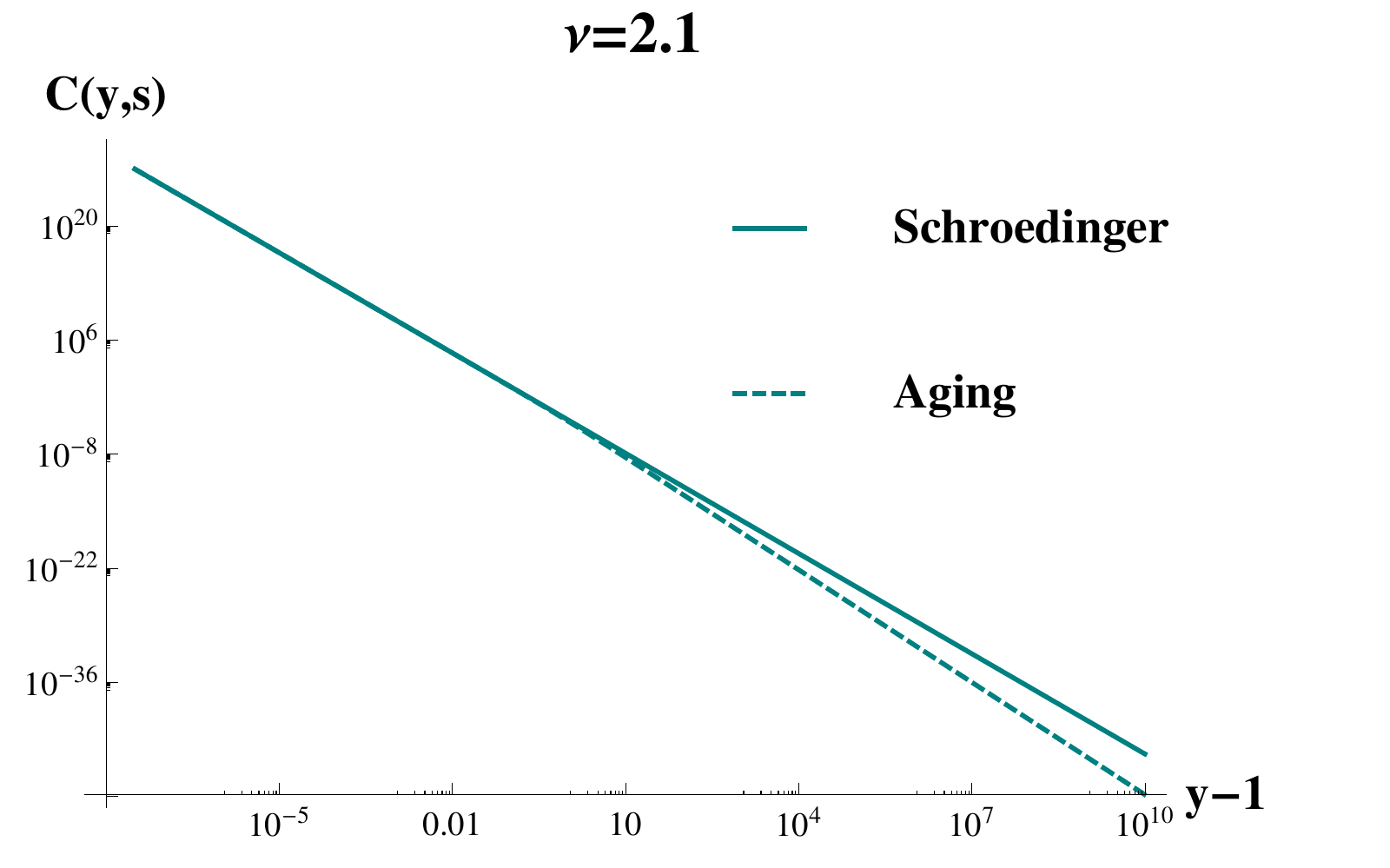}
 \end{tabular}
 \caption{\footnotesize Left: Log-log plot of the correlation function $C(y, s) $ vs. $y$ given in (\ref{TwoTimeCorr}),
 which shows a typical behavior of aging and is compared with the typical aging plot from numerical study, 
 {\it e.g.} fig. 1 of \cite{Henkel:2011NP}. Right: Log-log plot of $C(y, s) $ vs. $y-1$.
 }
\label{fig:OriginalLogLogCorreFunction}
\end{center}
\end{figure}
\vspace{-0.15in}

With these discussions in mind, let us investigate our correlation functions in detail.
First of all, it is natural to have $ \nu \geq 2$ in our construction. For this range,
we depict the log-log plot of the correlation
functions for the Schr\"odinger and aging cases together in figure \ref{fig:OriginalLogLogCorreFunction}.
We remind that the roughness $w (l, t)$ shows the similar behaviors as the square root of the correlation function,
$ C(l, t)^{1/2}$. In the left plot, we observe the typical aging behavior, which can be compared to fig. 1 of
\cite{Henkel:2011NP}. In the right plot, we put both the Schr\"odinger and aging correlation functions to
see clear differences.
Note that our aging correlation function (\ref{TwoTimeCorr}) changes its behavior around $\ln [y-1] \approx 0$,
while that of the Schr\"odinger does not. The latter represents the characteristic features of the
system with translational invariance.

\begin{figure}[!hb]
\begin{center}
\begin{tabular}{cc}
	 \includegraphics[width=0.5\textwidth]{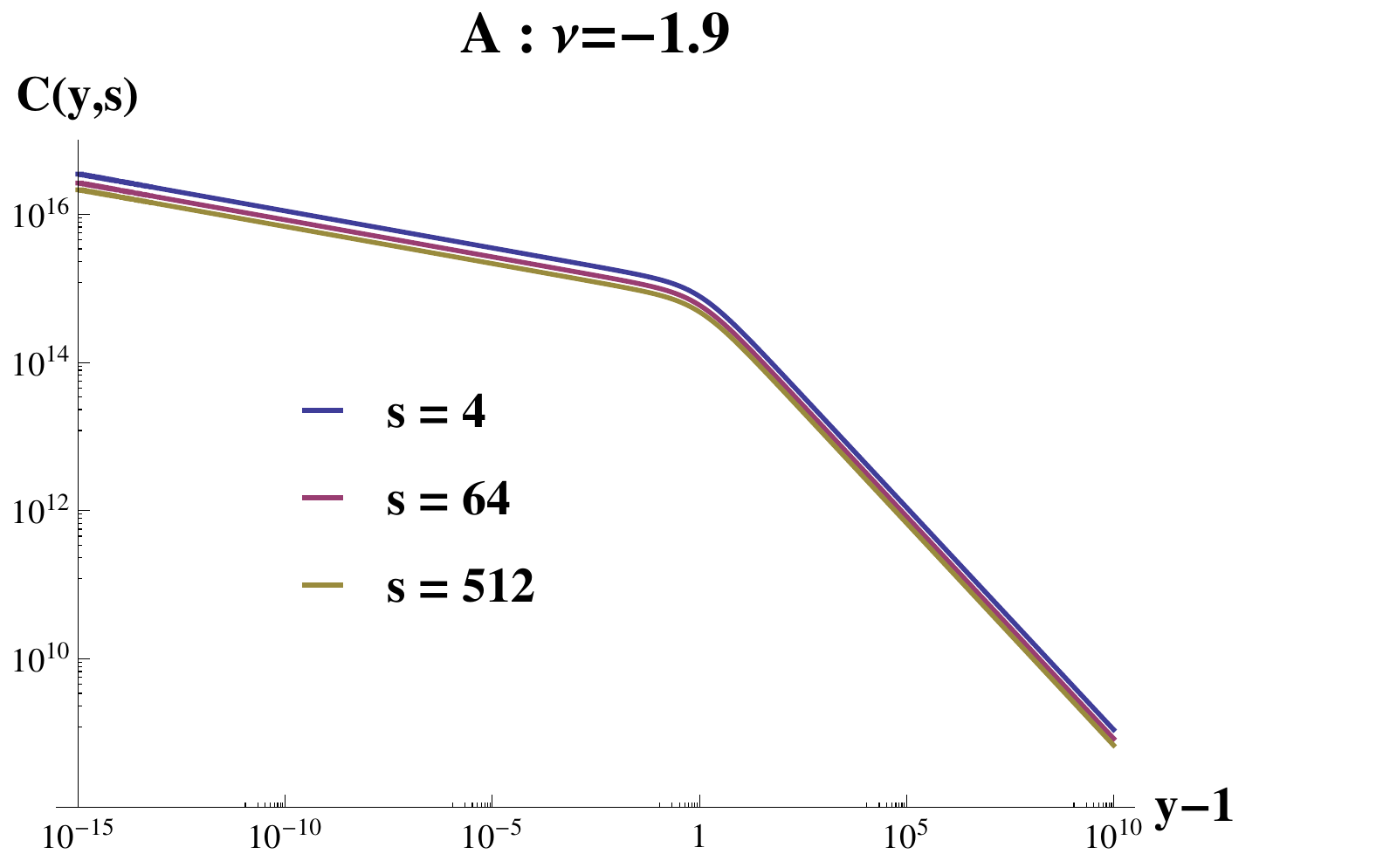}
	 \includegraphics[width=0.5\textwidth]{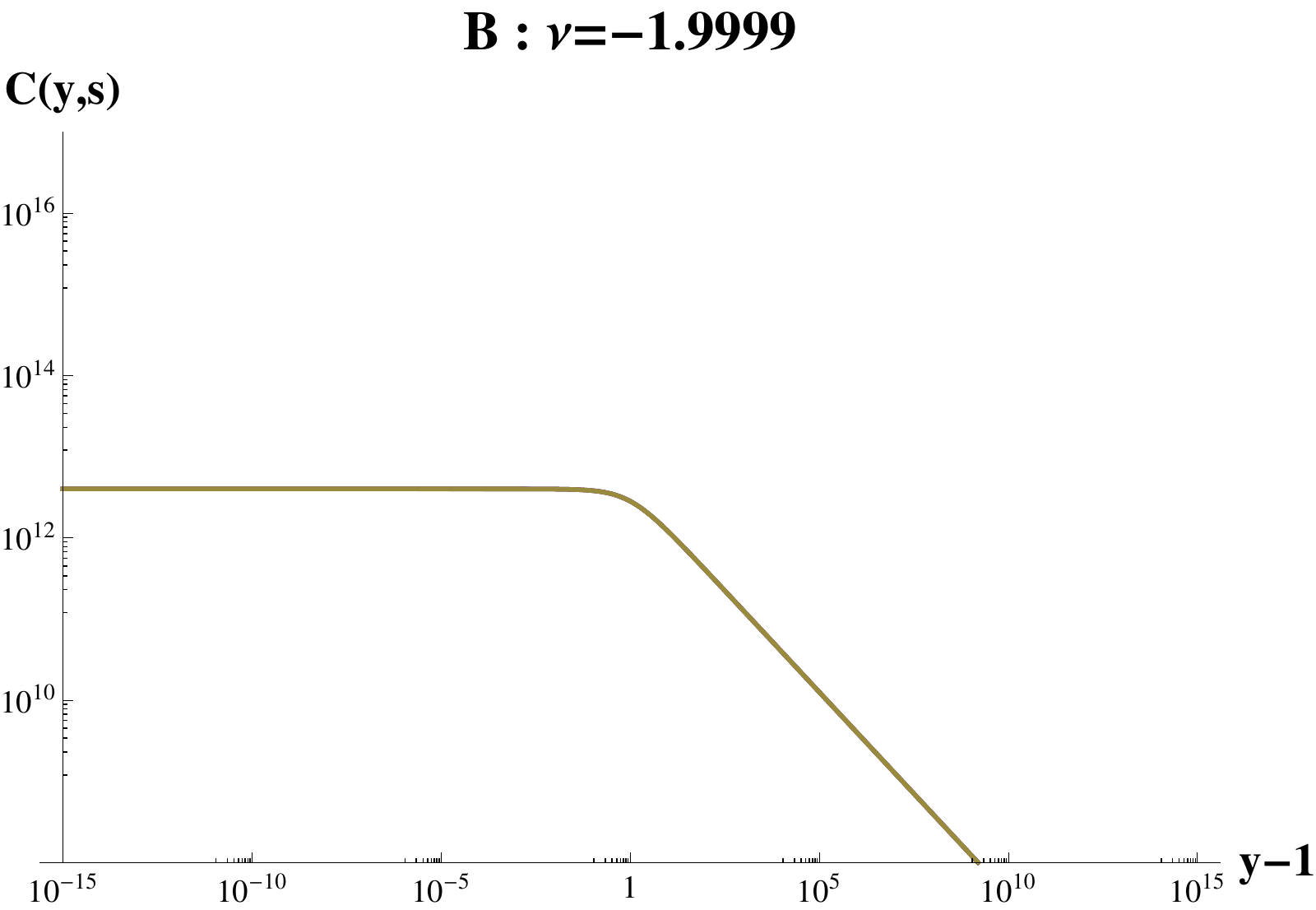}  \\
	 \includegraphics[width=0.5\textwidth]{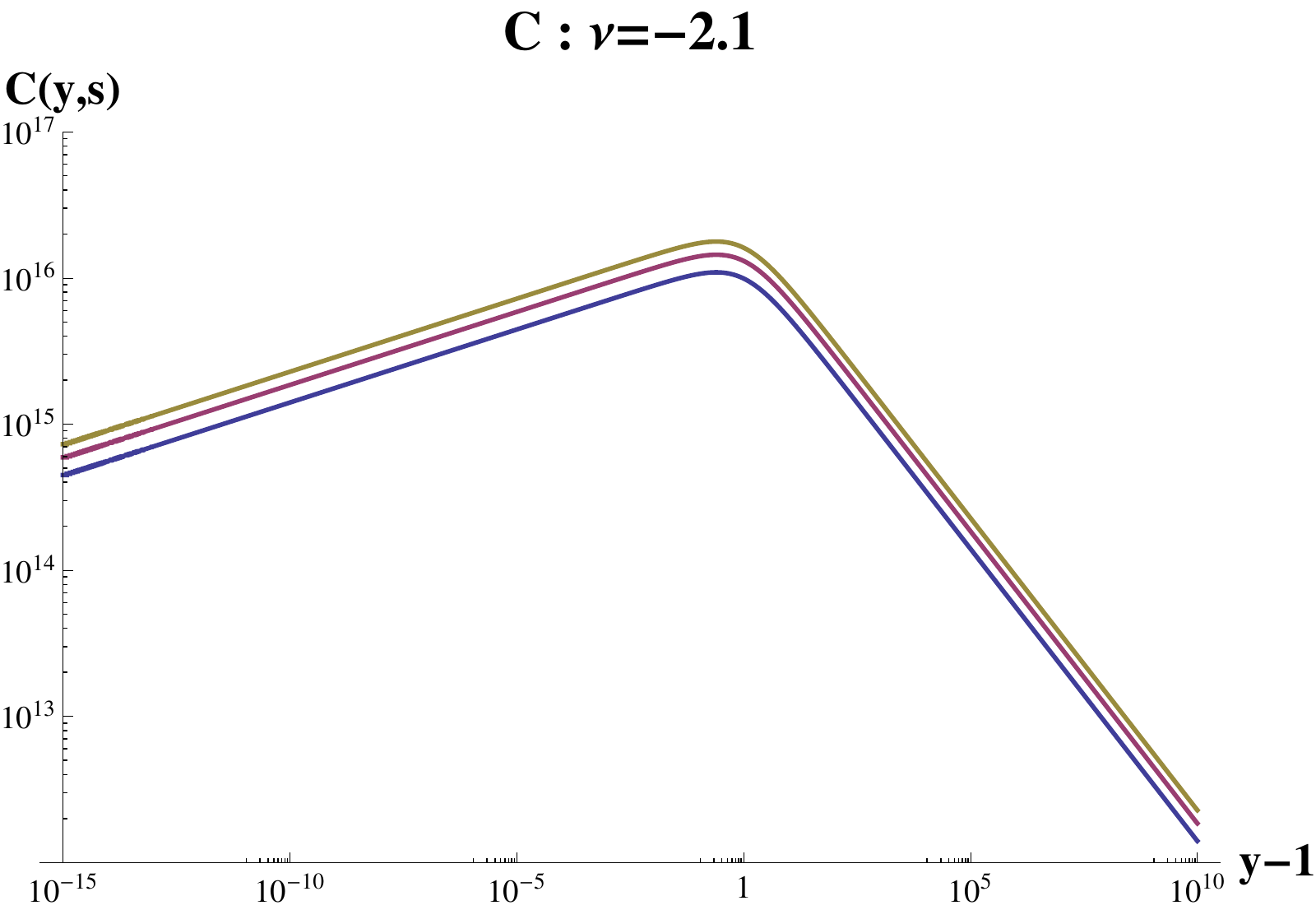}
	 \includegraphics[width=0.5\textwidth]{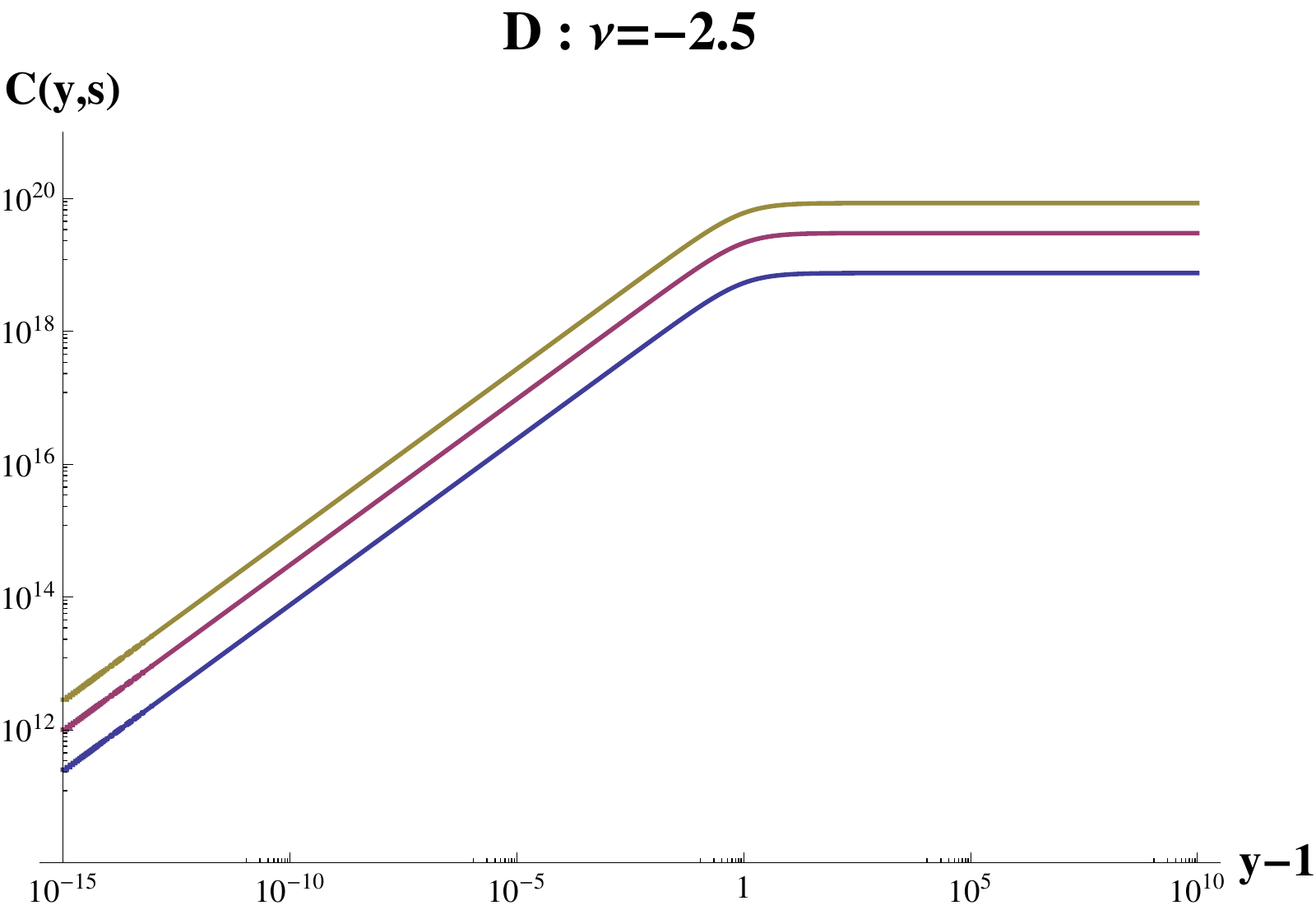}	
 \end{tabular}
 \caption{\footnotesize Log-log plot of the correlation function $C(x_2^+ , x_1^+) $.
 We clearly observe the growing or decaying patterns.
 The top left panel (A) shows only decaying behaviors, valid for $-2 < \nu $, while the bottom
 right panel (D) shows only growing ones valid for $\nu < -2-\frac{\alpha \mathcal M}{2}$.
 The bottom left panel (C) shows an interesting behavior, changing from growing to decaying,
 valid for $-2-\frac{\alpha \mathcal M}{2} < \nu < -2$. 
 Qualitative change is observed near $ \nu = -2 $ as shown in panel B.
 }
\label{fig:LogLogAgingPlotCorrelationFunction}
\end{center}
\end{figure}
\vspace{-0.2in}

It is interesting to extend our analysis to the range $ \nu < 2 $ from the phenomenological point of view. 
As mentioned above, there are qualitative changes at $\nu = -2 $ and 
$\nu = -2-\frac{\alpha \mathcal M}{2}$.
Let us concentrate on aging case, whose correlation function is fixed by our bulk geometry.
It is interesting to see the qualitative different behaviors of the correlation functions
by changing the values of  parameters, $ \nu$ and $\alpha \mathcal M $.
The typical behaviors are depicted in figure \ref{fig:LogLogAgingPlotCorrelationFunction}.
Their behavior for $\nu > -2 $, given in panel A of figure \ref{fig:LogLogAgingPlotCorrelationFunction},
is similar to that of the $\nu = 2.1 $ depicted in figure \ref{fig:OriginalLogLogCorreFunction}.
As $ \nu $ approaches the critical value $ -2 $, it becomes almost flat for small, positive $ \ln [y-1] $.
On the other hand, it shows the growing behavior for $\nu < -2-\frac{\alpha \mathcal M}{2}$.

Surprisingly, the bottom left panel (C), which is valid for the range
$-2-\frac{\alpha \mathcal M}{2} < \nu < -2$, shows a typical growing behavior for $\ln [y-1] \leq 0$,
while decaying (aging) behavior for $\ln [y-1] \geq 0$. We can see these changes because we choose
the horizontal axis as $ \ln [y-1] $. If we choose the horizontal axis as $ \ln [y] $ instead of $ \ln [y-1] $,
we only see the aging behaviors because the entire region $\ln [y-1] \leq 0$ shrinks to very small
part around the origin of $ \ln [y] $.

\subsection{Response function }   \label{sec:TwoTimeResponse}

Let us consider the two-time response function  $R(x_2^+ , x_1^+) = \langle {\phi}  {\phi} \rangle $,
which includes the logarithmic contribution obtained from (\ref{ZeroTCorrelatorAgingLog})
	\begin{align}   \label{TwoTimeLogResponseA}
		R(x_2^+ , x_1^+) &=  (x_2^+ - x_1^+)^{-2-\nu}
		\cdot \left( \frac{x_2^+}{x_1^+} \right)^{-\frac{\alpha \mathcal M}{2} }
		\left( R_0 - \ln [x_2^+ - x_1^+] \right)  \;,
	\end{align}
where $ R_0 = 1/\nu  + \ln [M u_B^2/2] -\frac{\Gamma (\nu)'}{\nu \Gamma (\nu)} $.
This is again the function of the waiting time and response time.
See some relevant explanations on them near the equation (\ref{TwoTimeCorr}).

\begin{figure}[!ht]
\begin{center}
\begin{tabular}{cc}
	 \includegraphics[width=0.6\textwidth]{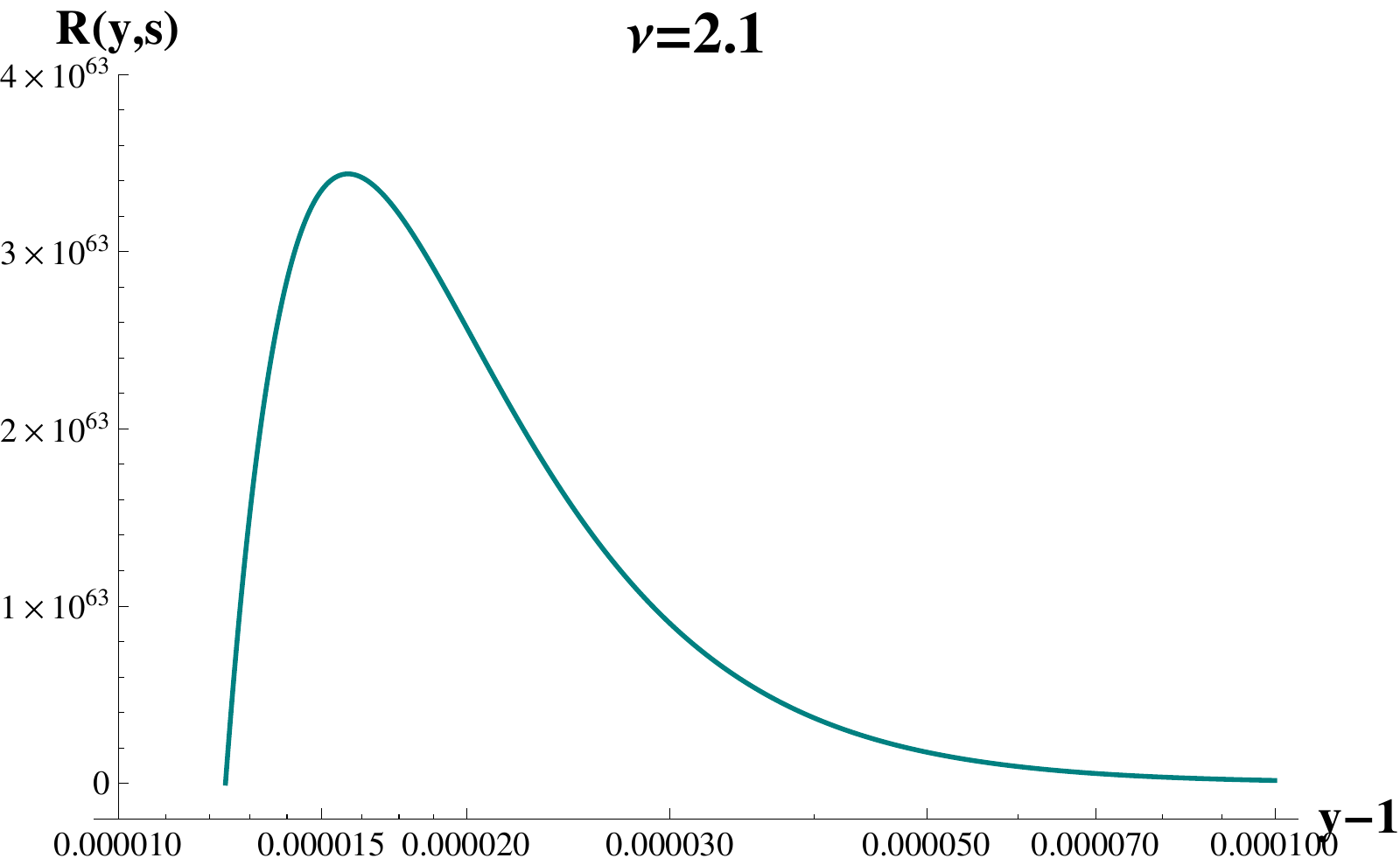}
 \end{tabular}
 \caption{\footnotesize Log-linear plot of the two-time response function with logarithmic extension 
 (\ref{TwoTimeLogResponse}) for $\nu=2.1 $ with fixed $\alpha \mathcal M = 1$. 
 This parameter range of $ \nu $ naturally arises from our holographic model.
 }
\label{fig:LogAgingPlotNu21}
\end{center}
\end{figure}
\vspace{-0.15in}

To analyze $R(x_2^+ , x_1^+) $, it is convenient to introduce a scaling time variable
$y = x_2^+ / x_1^+ $, which is the ratio between the response and waiting times.
Then
	\begin{align}  \label{TwoTimeLogResponse}
		R(s, y)	&= s^{-2 -\nu} y^{-2-\nu-\frac{\alpha \mathcal M}{2}}
		\left(1- \frac{1}{y} \right)^{-2 -\nu}
		\left( R_0 -  \ln s -  \ln y -  \ln [ 1- \frac{1}{y}]  \right) \;,
	\end{align}
where, once again, $s=x_1^+ $ is the waiting time, $1 \leq y = x_2^+ / x_1^+ \leq \infty$
and $\nu =\sqrt{4+m^2L^2 -\gamma \mathcal M^2}$.
Typical behaviors of response function with logarithmic extension from our model are depicted in
figure \ref{fig:LogAgingPlotNu21}. Similar behaviors of aging with logarithmic extension are noticed in
\cite{Henkel:2010hd}.

It is interesting to survey the general features of the response function $R(s, y) $, for the
different values of the parameter $ \nu $ as well as for the entire range of $ y$,
as a function of $y $.
Before proceed further, note that our correlation function has explicit
$\ln [1-1/y] $ and $\ln y $ dependence in addition to $\ln s $
as a consequence of the logarithmic extension.
One might wonder the effect of $\ln y $ which has the divergent behavior for large $y $.
It turns out this depends crucially on the value of the parameter $\nu $, which is
explained at the end of this section.

\begin{figure}[!ht]
\begin{center}
\begin{tabular}{cc}
	 \includegraphics[width=0.5\textwidth]{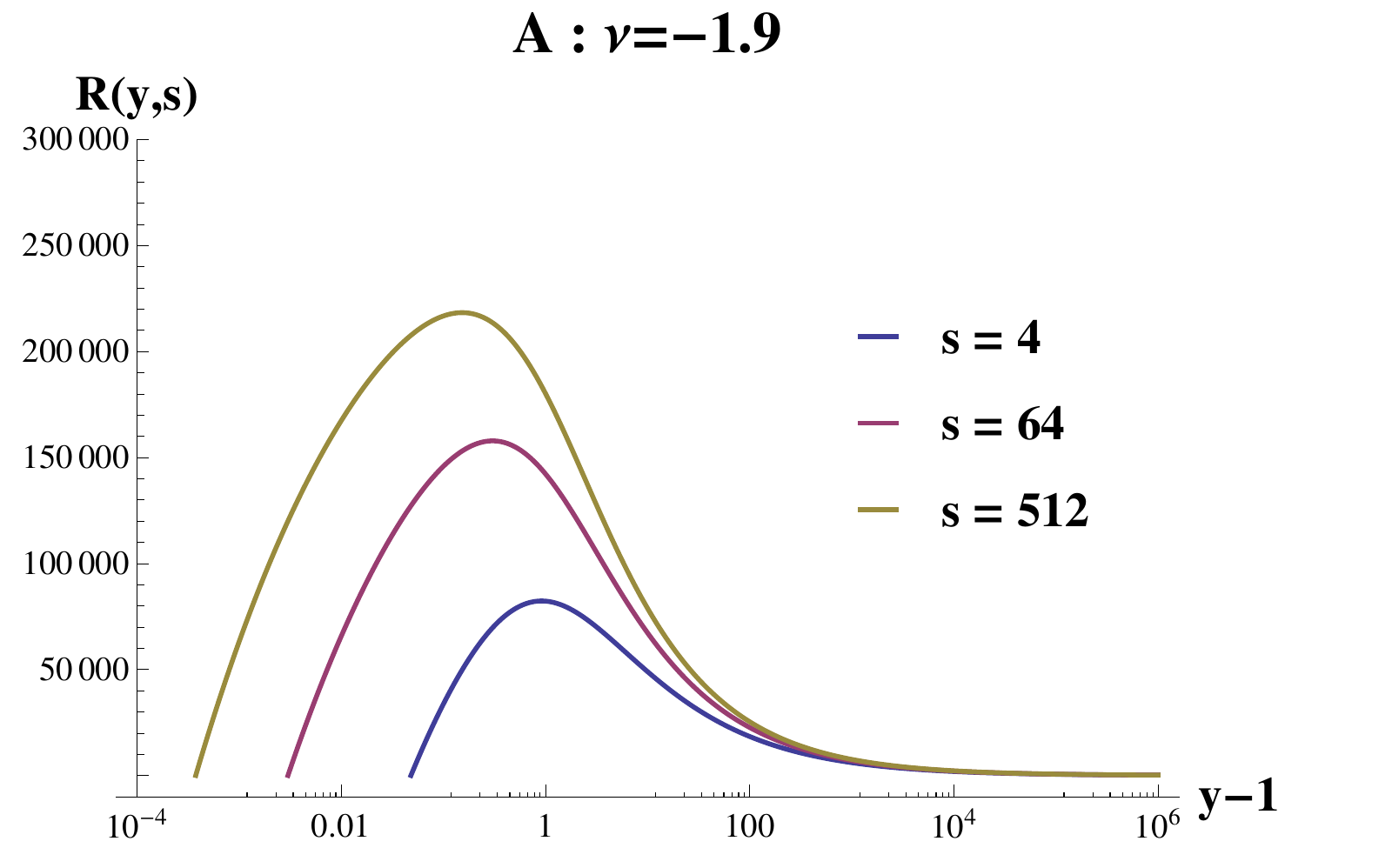}
	 \includegraphics[width=0.5\textwidth]{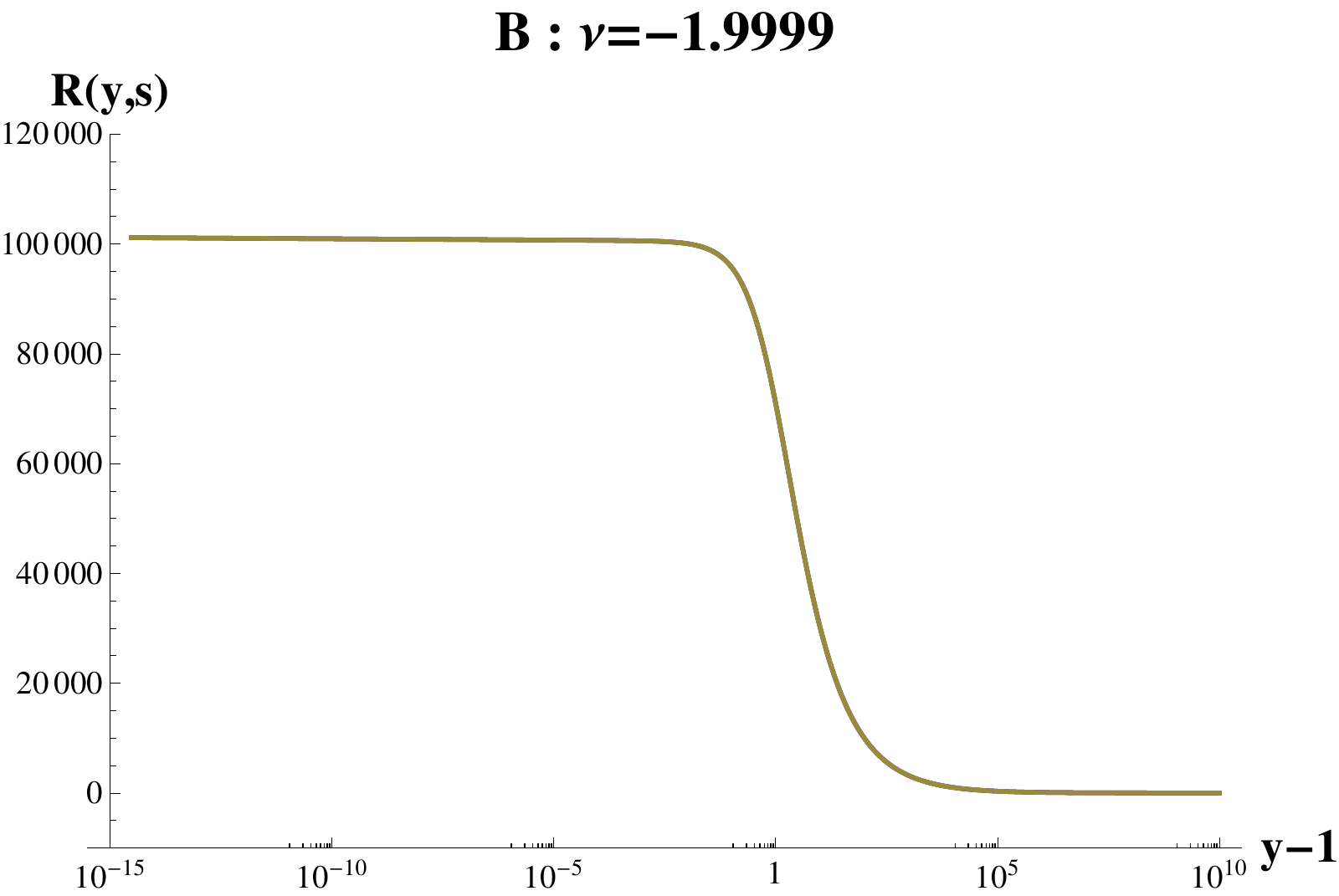} \\
	 \includegraphics[width=0.5\textwidth]{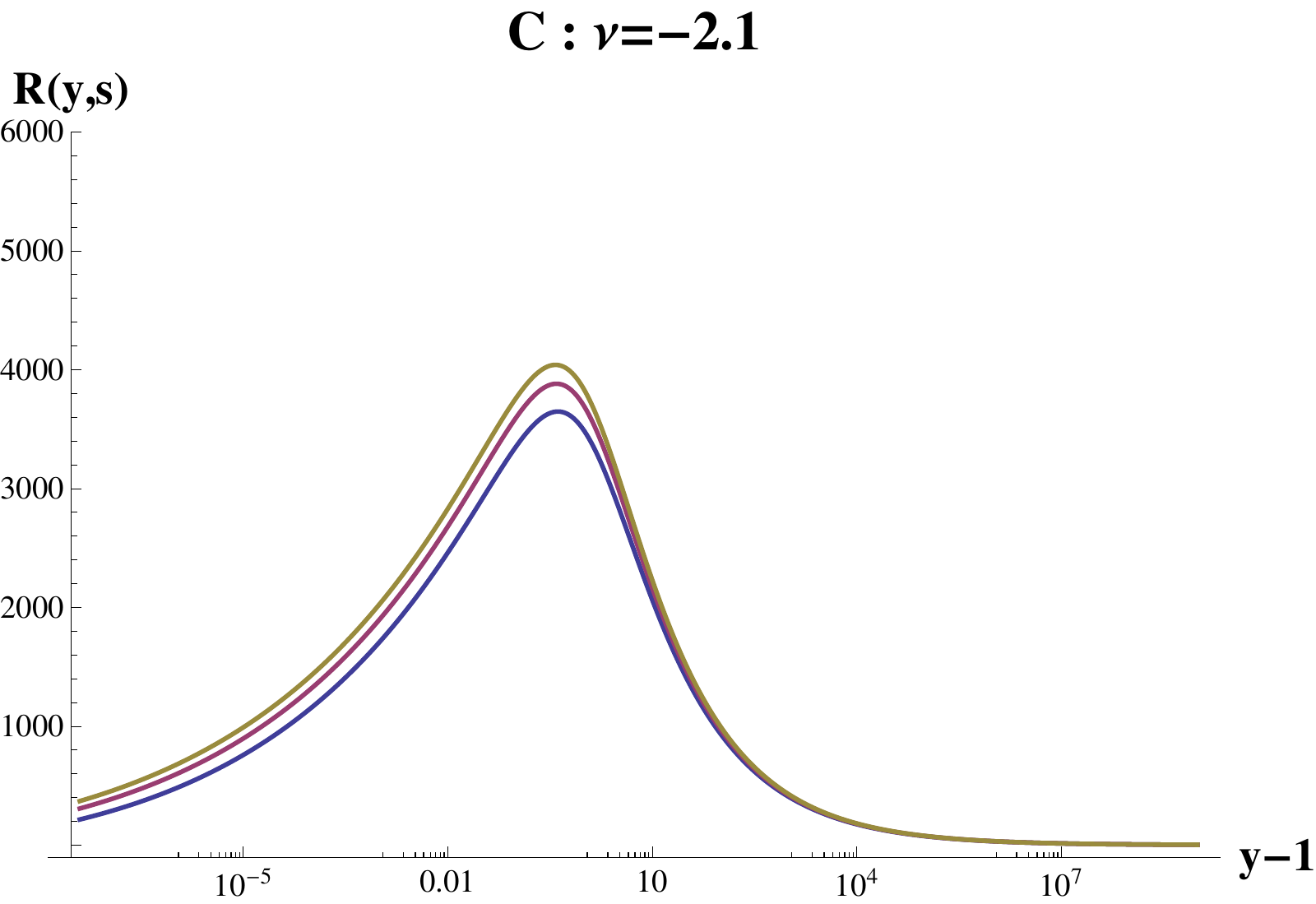}
	 \includegraphics[width=0.5\textwidth]{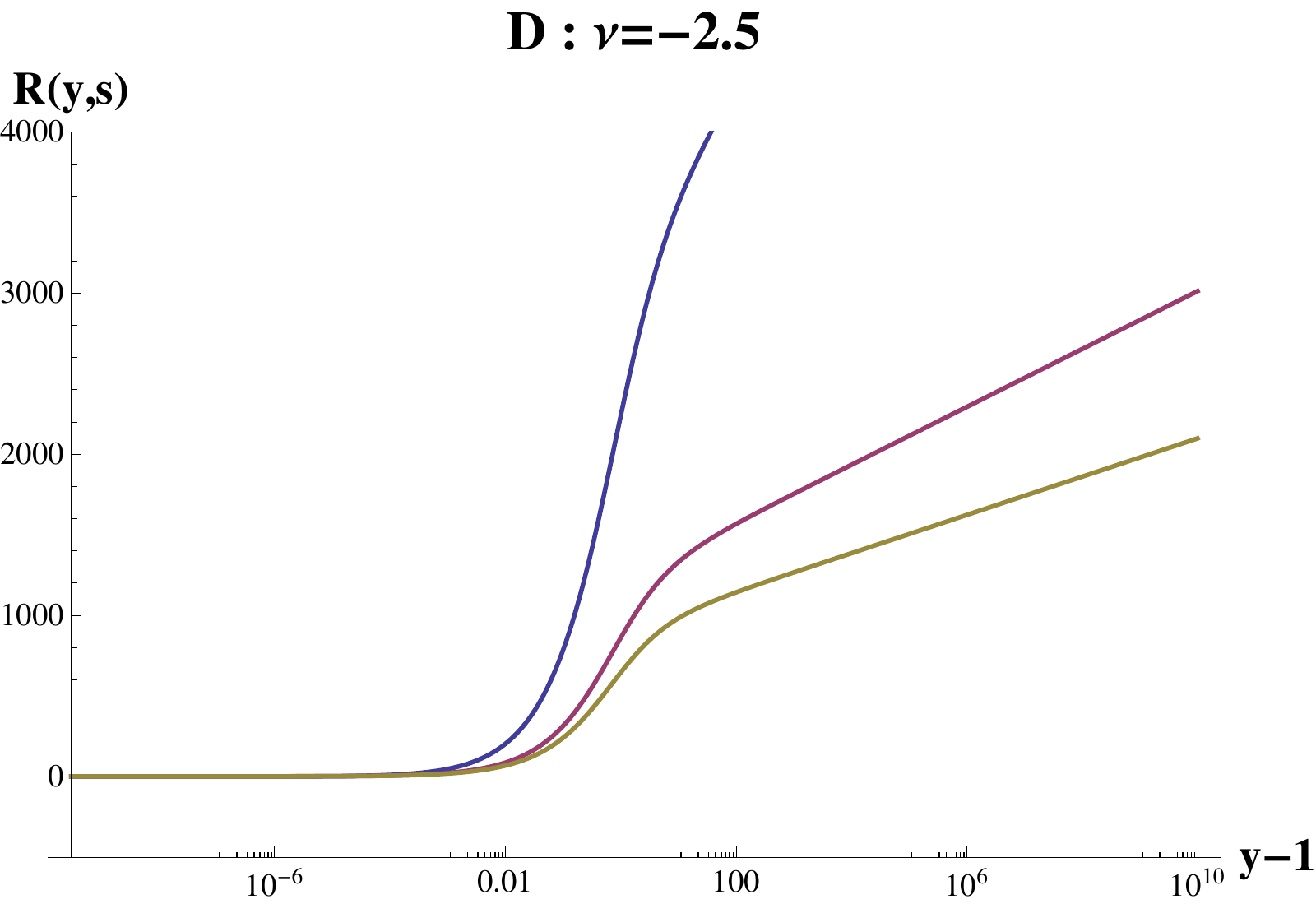}	
 \end{tabular}
 \caption{\footnotesize
 Plot of the two-time response function with logarithmic extension (\ref{TwoTimeLogResponse}) 
 for four different values of $\nu $ with fixed $\alpha \mathcal M = 1$. 
 The horizontal axis is $y-1$. Each panel has three graphs with different waiting times.
 They are normalized to be unity at $y=10^{10} $ to show the scaling behavior, and thus the effect
 of the factor $ s^{-2 -\nu}$ is washed out. The remaining effects of the waiting time are from
 $\ln s $ in (\ref{TwoTimeLogResponse}).
 The top left panel A is typical for $ \nu > -2 $, which converges to $0 $ for large $y $ and diverges
 toward the negative infinity for small $y $. The panel B is for near $ \nu = -2 $, which shows the transition
 behavior of $R(s, y)$.  Panel C shows a typical behavior for
 $-2-\frac{\alpha \mathcal M}{2} < \nu < -2 $, which nicely converges for both large and small $y $.
 Panel D is for $ \nu = -2-\frac{\alpha \mathcal M}{2} $, which shows a diverging behavior for large $y $ and
 thus is normalized at $ y=10^{-10} $ instead.
 }
\label{fig:LogAgingPlot}
\end{center}
\end{figure}
\vspace{-0.15in}

There exist essentially four different
types of qualitatively different behaviors, two from large $y $ and two from small $y $.
At large $y $, $ 1-1/y $ becomes a unity. Thus the behavior of $R(s, y) $ is controlled by
the combination
	\begin{align}
		y^{-2-\nu-\frac{\alpha \mathcal M}{2}} \ln y \;,
	\end{align}
which converges for $ \nu > -2-\frac{\alpha \mathcal M}{2} $ and diverges otherwise.
At small $y \rightarrow 1 $, the behavior of $R(s, y) $ is controlled by
the combination
	\begin{align}
		\left(1- 1/y \right)^{-2 -\nu} \ln [ 1- 1/y]   \;,
	\end{align}
which converges for $ \nu < -2 $ and diverges otherwise.
From these behaviors, we conclude that the two-time correlation function converges both for
small and large $y $ for the range 
	\begin{align}
		-2- \alpha \mathcal M / 2 < \nu < -2 \;.
	\end{align}
This case is depicted in the bottom left panel (C) of figure \ref{fig:LogAgingPlot}.

\begin{figure}[!b]
\begin{center}
\begin{tabular}{cc}
	 \includegraphics[width=0.5\textwidth]{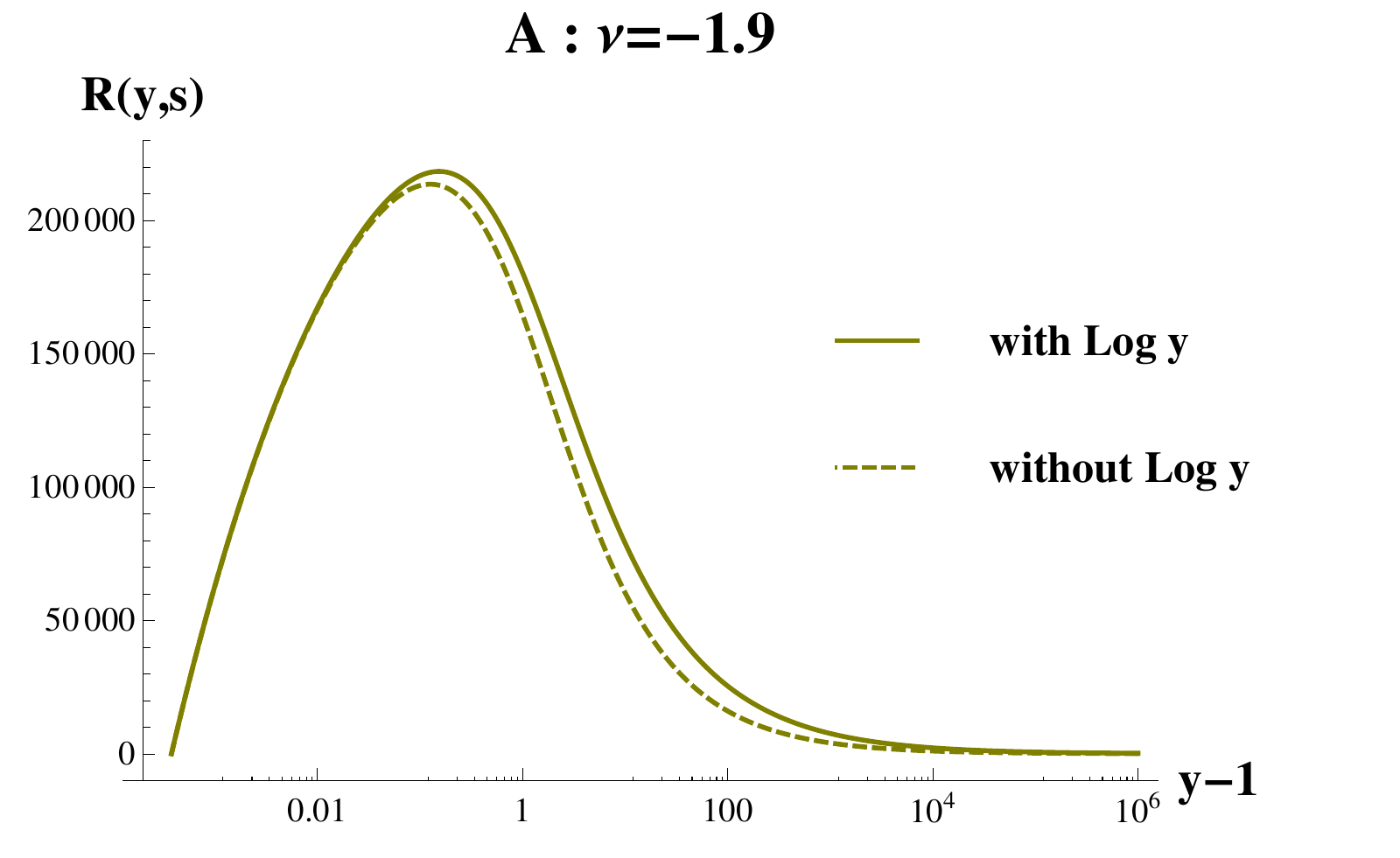}
	 \includegraphics[width=0.5\textwidth]{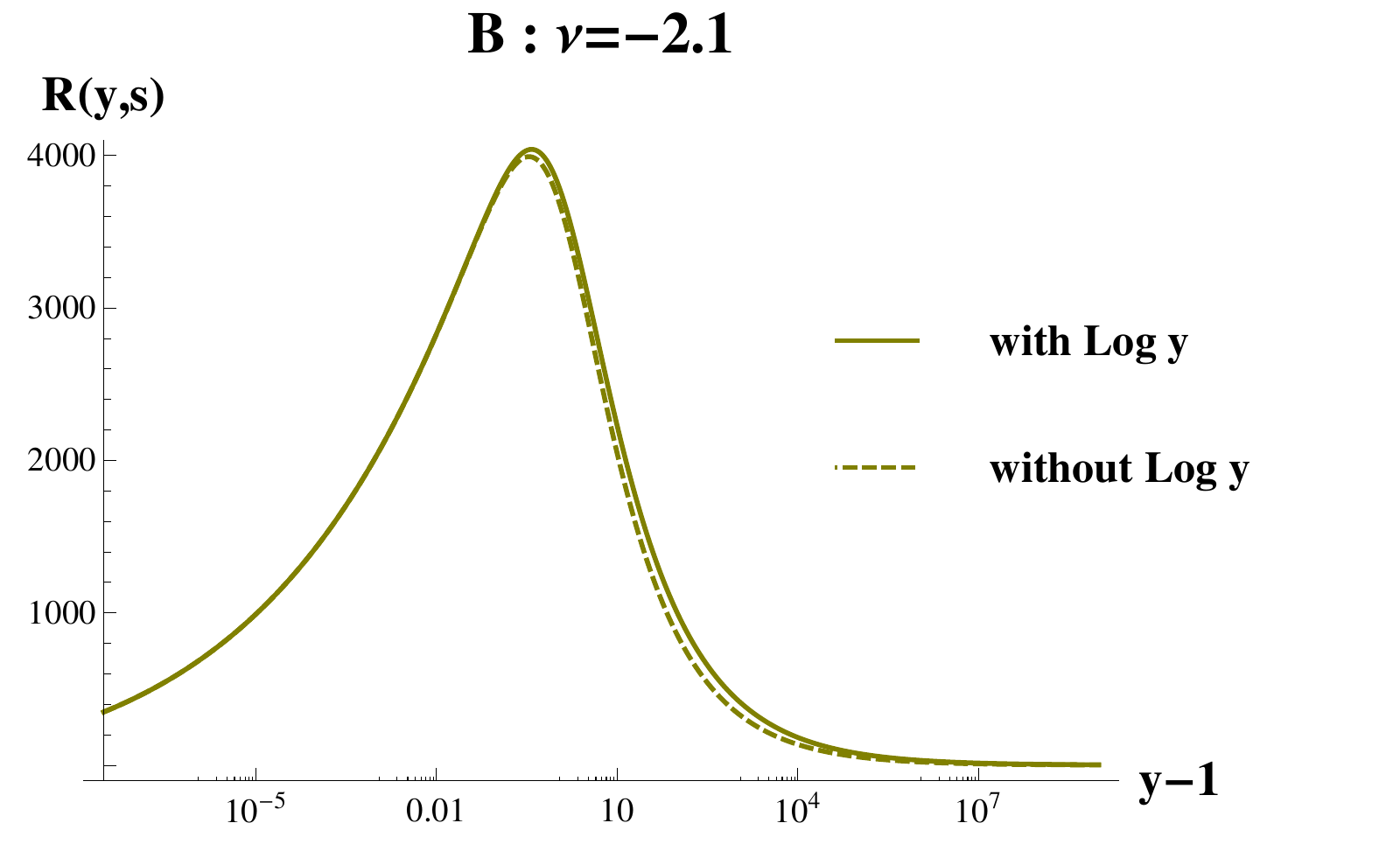}
 \end{tabular}
 \caption{\footnotesize
 Plots for the two-time response function (\ref{TwoTimeLogResponse}) against $R_L (s, y) $
 without the $\ln y $ term given in (\ref{TwoTimeLogCorrWithoutLog}) for two different values of $\nu $.
 They are normalized by the value of (\ref{TwoTimeLogResponse}) at $y=10^{10} $ to show clear differences.
 The left panel is a typical case for $ \nu > -2 $, while the right panel is for
 $-2-\frac{\alpha \mathcal M}{2} < \nu < -2 $. For both cases, the effect of $\ln y $ is negligible
 for small $y $, while there are visible differences for large $y $.
 Even taking into account of all these differences, the qualitative features remain intact for
 $\nu > -2-\frac{\alpha \mathcal M}{2} $, and thus physical applications would not be affected much.
 }
\label{fig:LogAgingPlotWithoutLog}
\end{center}
\end{figure}

Let us comment the effect of $\ln y $, which would diverge for large $ y $, as we mention above.
Certainly, there exist a divergent behavior at large $y $ for the range $\nu < -2-\frac{\alpha \mathcal M}{2}$
as depicted in the bottom right panel (D) of figure \ref{fig:LogAgingPlot}.
For the other range, the effect of the $\ln y $ does not produce qualitatively different behaviors
in the response function. We investigate this issue by plotting the correlation function without
the $\ln y $ term
	\begin{align}   \label{TwoTimeLogCorrWithoutLog}
		R_L (s, y)
		&= s^{-2 -\nu} y^{-2-\nu-\frac{\alpha \mathcal M}{2}}
		\left(1- \frac{1}{y} \right)^{-2 -\nu}
		\left( C_0 - \ln s  - \ln [ 1- \frac{1}{y}]  \right) \;,
	\end{align}
against the two-time response function $R(s, y) $ given in (\ref{TwoTimeLogResponse}).
The plot is depicted in figure \ref{fig:LogAgingPlotWithoutLog}.
As expected, there are some visible differences for large $y $. But the differences do not
change the qualitative features of the plot, and accordingly physical applications would not
be affected much by the effect of $\ln y $.

With these observations, we go on to investigate the possibility that our system might be
relevant for the physics of logarithmic extension of aging in the following section.

\subsection{Logarithmic extension of aging in KPZ class}

There have been various attempts to understand the logarithmic extension
of the non-relativistic AdS/CFT \cite{Bergshoeff:2011xy}.
Recently, the logarithmic corrections to the aging correlation and response functions are introduced
to investigate their connections to the aging phenomena in field theory context, where
their dynamical scaling is characterized by the aging exponents \cite{Henkel:2011NP}\cite{Henkel:2010hd}.
Specifically the authors of \cite{Henkel:2011NP} study the aging phenomena of the surface growth
processes described by (1+1)-dimensional Kardar-Parisi-Zhang equation \cite{KPZ}.
The associated dynamical exponent is $3/2 $.

While we are very interested in providing detailed analysis of our model,
quantitative comparisons are not feasible due to several reasons at this point: our model is
designed for $2+1 $ dimensional dual field theory with dynamical exponent $z=2 $ exactly
at the critical point.%
\footnote{The case with $ z=3/2 $ for arbitrary $d+1 $ dimensions has been considered
in some detail \cite{Kim:2012nb}. There, correlation function without logarithmic extension
is obtained for the special case, without full generality due to the technical difficulties.
\label{footnote:ZThreeHalf}
}
We already presented our qualitative results with the logarithmic contributions in the previous section.
Let us comment the similarities and differences compared with \cite{Henkel:2011NP}\cite{Henkel:2010hd}.
Especially, the equation (4.3) of \cite{Henkel:2010hd} (or similarly equation (10) of \cite{Henkel:2011NP})
reads (using the same notation as in (\ref{TwoTimeLogResponse}))
	\begin{align}   \label{FTresponseFunction}
		R(s,y) &=  s^{-1 -a} y^{-\lambda_R /z} \left(1- \frac{1}{y} \right)^{-1-a'}  \\
		&\times \left( h_0 - g_{12,0} \tilde \xi' \ln [ 1- \frac{1}{y}] -\frac{1}{2} f_{0}
		\tilde \xi'^2\ln^2 [ 1- \frac{1}{y}]
		- g_{21,0} \xi ' \ln [ y- 1] + \frac{1}{2} f_{0} \xi'^2 \ln^2 [ y-1] \right) \;, \nonumber
	\end{align}
where the parameters $g_{12,0}, g_{21,0}, \tilde \xi', \xi', f_{0}$ come from the logarithmic extension
in the field theory side. From the phenomenological input \cite{Henkel:2010hd}
``the parenthesis becomes essentially constant for sufficiently large $ y $,''
the condition $ \xi ' =0 $ is imposed to remove the last two terms.
The other parameters in (\ref{FTresponseFunction}) are fixed to the data.
Compared to our result (\ref{TwoTimeLogResponse}), we can identify the various exponents
	\begin{align}
		\nu = a -1 \;, \qquad
		\frac{\alpha \mathcal M}{2} = \frac{\lambda_R}{z} -1 - a \;,
	\end{align}
along with the condition $ a' =a $, which is a restricted case.

There are also three notable differences in the parenthesis.
First, we note the existence of the term $ \ln y $, which might spoil the large $ y $ behavior.
As we discussed in the previous section \S \ref{sec:TwoTimeResponse},
the response function (\ref{TwoTimeLogResponse}) converges at large $ y $ for the parameter range
$ \nu > -2 -\frac{\alpha \mathcal M}{2} $. Thus there exist large parameter spaces we can reliably
describe the desired properties of the response function.
Second, we also have an explicit dependence of the waiting time $ s $. Even after the normalization
with $ s^{2 +\nu} C(s, y) $ to remove the dependence on the waiting time, there exist another source
of the contribution on $s $, which is the important consequence of the broken time translational invariance and 
might provide interesting physical properties.  
Third, the correlation function (\ref{TwoTimeLogResponse}) does not
have $ \ln^2 [ y-1] $ term. Seemingly this is a small correction and thus does not give qualitatively
different features.%
\footnote{In the context of the log CFT and its geometric realizations, an extension to the tri-critical
gravity is shown to produce the $ \ln^2 [ y-1] $ contributions \cite{Bergshoeff:2012ev}.
Thus it is reasonable to expect that
similar generalization would give the term in the case of Schr\"odinger holography, too. 
We thank to Y. S. Myung for the comment on this issue.
}

We don't claim that the response function (\ref{TwoTimeLogResponse}) {\it quantitatively}
describes the scaling properties of the growth phenomena for several reasons, including the
different spatial dimensions, different dynamical exponents and different terms present.
As far as we understand, there exist tensions between the experimentally realized growth
and the theoretically investigated aging phenomena for the KPZ universality class.
Our model provides several qualitatively distinctive behaviors of the correlation and response functions.
In a particular case, the two-time correlation and
response functions reveal an initial growth behavior as well as the slow power law decay,
typical for aging, at later times for some parameter ranges.
Further investigations along this line seem to be very interesting.
In particular, studying the $1+1 $ dimensional aging system with logarithmic extension for $z=3/2 $
would be very interesting for obvious reasons.

\section{Conclusion}   \label{sec:Conclusion}

Physics out of equilibrium is challenging, yet very important.
The aging system in holography as well as in field theory provides a simple framework describing
time dependent, thus far from equilibrium, physical systems with local scaling symmetry.
It has been known that corresponding correlation functions reveal a slow dynamics with power law decay,
broken time translational invariance and dynamical scaling in the aging regime, where waiting time $s $,
response time $ t $ and the scaling time $y=t/s $ are large compared to
the microscopic time scale of the system considering \cite{HenkelBook2}\cite{Hyun:2011qj}.

1+1 dimensional Kardar-Parisi-Zhang (KPZ) universality class \cite{KPZ} was realized recently
in growing interfaces of turbulent liquid crystals,
which is described by the dynamical exponent $ z=3/2$, growth exponent $\mathrm a =1/2 $ and roughness exponent
$\mathrm b=1/3 $ \cite{TakeuchiPRL}\cite{TakeuchiSciRep}.
Subsequently, aging properties of the same KPZ universality class in 1+1 dimension are studied
in \cite{Henkel:2011NP} with logarithmic extension of the correlation and response functions,
where the critical exponents are further extended. These results serve as our motivation.
We extensively study our holographic model to reveal the qualitative features of correlation and
response functions and to seek the connection to the KPZ universality class.

In this paper, we investigate the entire range of the scaling time $ y $ without assuming the aging regime,
especially the early time profiles of the two-time correlation and response functions.
We extensively study the typical patterns of them for the different parameter range with phenomenological
point of view.
In physical terms, the roughness of the growing interface and the square root of the
height-difference correlation function show similar behaviors in the KPZ universality class.
There are three qualitatively distinct roughness behaviors in log-log plots summarized in figure
\ref{fig:LogLogAgingPlotCorrelationFunction}:
i) slowly decreasing in early times and continue decreasing faster in later times,
ii) increasing in early times and change into decreasing in later times and
iii) fast increasing in early times and continue increasing slowly in later times.
In particular, the second case shows an initial growing behavior, typical for the growth,
followed by power law decaying behavior, typical for the aging for the parameter range
$-2-\frac{\alpha \mathcal M}{2} < \nu < -2 $, which is unexpected and interesting.
For the Schr\"odinger case, there are only two exclusive options, either growing or aging, in contrast.

It will be interesting to construct the corresponding holographic
model to check whether the holography can reproduce these critical exponents and
provide more useful information on this universality class.
To do so, it is required to extend our construction to non-conformal cases for $ z=3/2 $
in $ 1+1 $ dimensions, where analytic results are not available yet.

\section*{Acknowledgments}

We would like to thank to C. Hoyos, E. Kiritsis, B.-H. Lee, Y. S. Myung, Y. Oz, C. Park, 
S.-J. Sin, C. Sonnenschein and S.-H. Yi for discussions, comments and correspondences. 
SH is supported in part by the National Research Foundation of Korea(NRF) grant funded 
by the Korea government(MEST) with the grant number 2009-0074518 and the grant number 2012046278. 
SH and JJ are supported by the National Research Foundation of Korea (NRF) grant funded 
by the Korea government(MEST) through the Center for Quantum Spacetime(CQUeST) of Sogang University 
with grant number 2005-0049409.
BSK is grateful to the members of the CQUeST for their warm hospitality, 
especially to B.-H. Lee. This work has been finished during his visit to CQUeST.

\appendix

\section*{Appendix}

\section{Explicit evaluations of the logarithmic extension}   \label{sec:LogEvaluation}

In this section, we show the details of the inverse Fourier transform of ${\cal F}_2$.
The equivalent and simple derivations are presented in sections \S \ref{sec:SchrLCFT}
and \S \ref{sec:AgingLCFT} by taking the derivative of ${\cal F}_1$ with respect to
the parameter $ \nu $.

From (\ref{F1F2}), the non-trivial contribution of ${\cal F}_{2}$ can be expressed as
\begin{align}
        {\cal{F}}_{2} &= \frac{L^{3}}{u^{3}} c_{22} u^{2} K_{\nu} (\tilde{q} u) \frac{\partial}{\partial u}
       \left( \frac{1}{2\nu} \frac{d}{d\nu} \left(  c_{22} u^{2} K_{\nu} (q u)\right) \right)\Big|_{u_{B}}\\
       & \approx   \frac{L^{3}}{2\nu u_{B}^{4} } q^{2\nu} \frac{d}{d \nu}
       \left( 2\nu \frac{ \Gamma(-\nu)}{\Gamma(\nu)}\left( \frac{u_{B} }{2} \right)^{2\nu} \right)
       +  \frac{L^{3}\Gamma(-\nu)}{u_{B}^{4}\Gamma(\nu)} \left( \frac{u_{B} }{2} \right)^{2\nu} q^{2\nu} \ln q ^{2} \;,
       \label{F2function}
    \end{align}
where we expand the expression for small $ u_B $.
The inverse Fourier transform (IFT) of $q^{2\nu}$ in the first term is the same as ${\cal F}_{1}$.
Thus we concentrate on IFT of $q^{2\nu} \ln q^{2}$ in the second term.

\begin{figure}[!h]
\begin{center}
\begin{tabular}{c}
	 \includegraphics[width=0.45\textwidth]{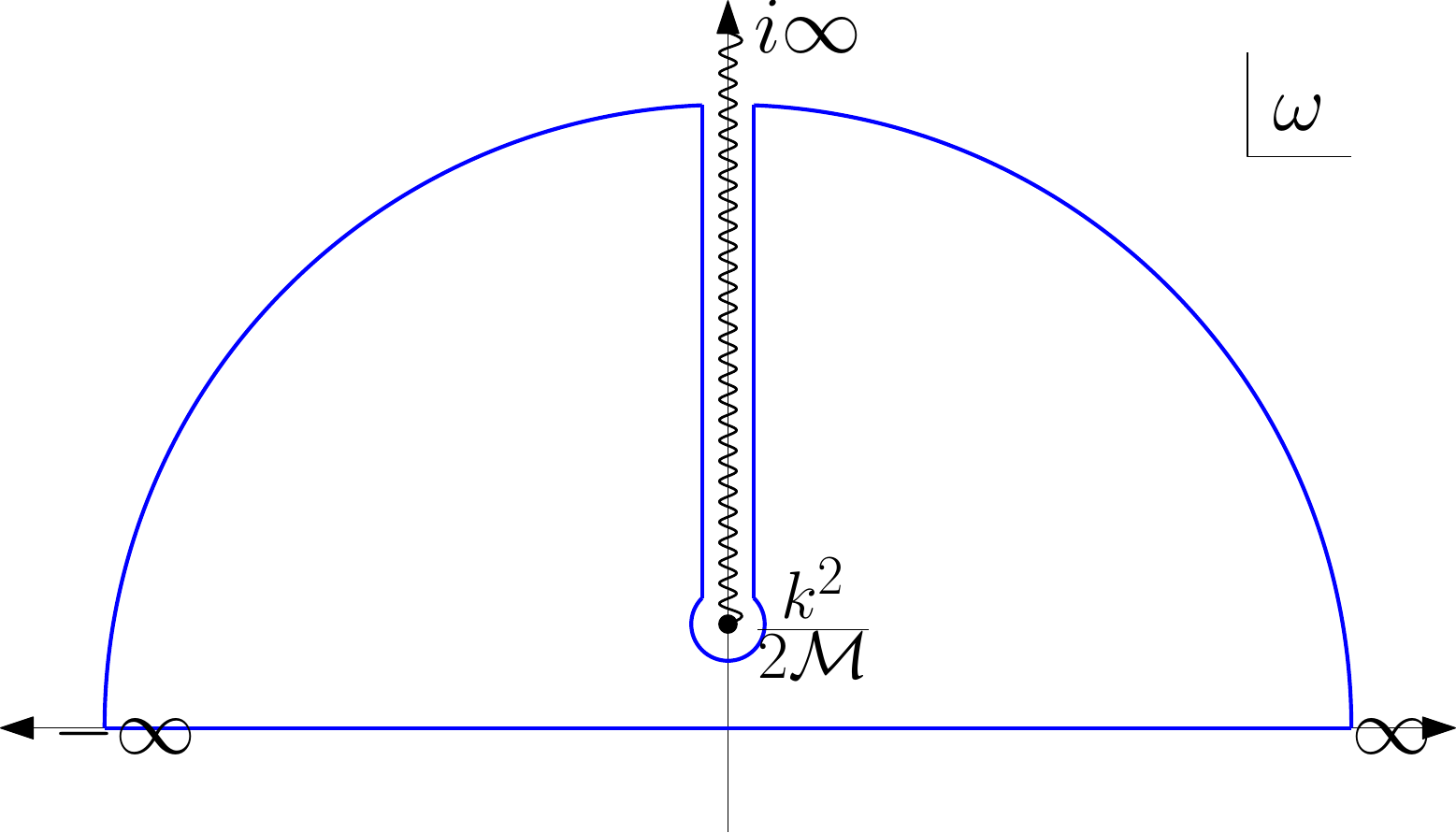} \qquad \qquad
	 \includegraphics[width=0.35\textwidth]{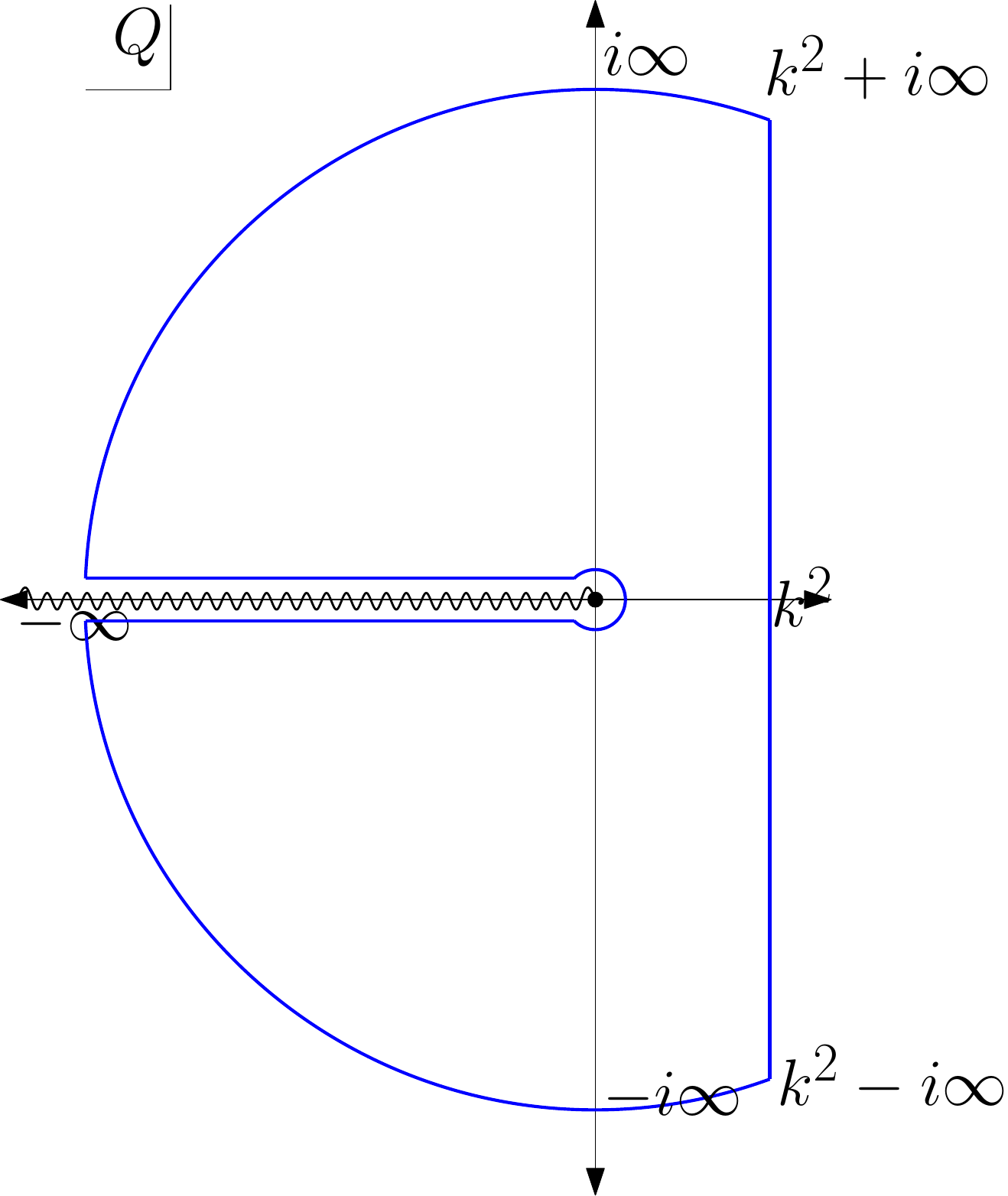}
 \end{tabular}
 \caption{Left : non-trivial contour in the complex $\omega$ plane for positive ${\cal M}$.
 Right : contour in the complex $Q$ plane. The integrals at infinity and around origin vanish.}
\label{fig:ContOmeg}
\end{center}
\end{figure}

By defining $Q\equiv q^{2} = \vec{k}^{2} + 2 i {\cal M} \omega$, we compute IFT
with respect to $\omega$ as
\begin{align}
    &\int_{-\infty}^{\infty} e^{ i \omega t } q^{2\nu} \ln{q^{2}}\, d\omega
     =\int_{-\infty}^{\infty} e^{ i \omega t } Q^{\nu} \ln{Q}\, d\omega   \nonumber\\
    &\qquad = - \frac{i}{2 \left| \cal M \right| } \exp \left( -\frac{k^{2}}{2 \cal M} t \right)
    \int_{ k^{2}- i \cdot \infty }^{k^{2} + i \cdot \infty} \exp \left( \frac{Q}{2 \cal M} t \right)
    Q^{\nu} \ln Q  d Q \;.
    \label{eq:fourierQ}
\end{align}
For ${\mathcal M} > 0$, a branch point is located at $Q_{0} =0$,
while $\omega_{0} = \frac{ i k^{2}}{2\cal M}$ is in the upper half plane.
To get a nonvanishing contribution for positive $ t $, we choose our contour to the
upper half part in complex $ \omega $ plane. This is depicted in the left side
in figure \ref{fig:ContOmeg}. Similarly, in the $Q$ plane, the contour is left half plane
of the line ${\textrm{Re}}[ Q]=k^{2}$ and branch cut lies in the negative direction from the origin
($Q=0$), which is depicted in the right part of the figure \ref{fig:ContOmeg}.

From the contour integral in complex $Q $ plane, the nonvanishing contributions come from two terms
	\begin{align}
	&\int_{-\infty+ i \epsilon}^{0+i \epsilon} \exp \left( \frac{Q}{2 \cal M} t \right) Q^{\nu} \ln Q \, d Q +
	\int_{0- i \epsilon}^{-\infty-i \epsilon} \exp \left( \frac{Q}{2 \cal M} t \right) Q^{\nu} \ln Q \, d Q \;.
	\end{align}
In the first term we identify $ Q=|Q|e^{i \pi} $, while $ Q=|Q|e^{-i \pi} $ in the second term. Thus we get
	\begin{align}
	2\pi i \cos( \pi \nu) \int_{0}^{\infty} \exp \left( - \frac{t}{2 \cal M} Q \right) Q^{\nu} \,d Q + 2 i \sin(\pi \nu) \int_{0}^{\infty} \exp \left( - \frac{t}{2 \cal M} Q \right) Q^{\nu} \ln Q\, d Q \;,
	\end{align}
which can be evaluated to be the following using the definition of the $ \Gamma $ function 	
	\begin{align}
	\Gamma ( \nu +1) \left( \frac{ 2 \cal M }{t}\right)^{\nu+1} \left[ 2\pi i \cos( \pi \nu)
	+ 2 i \sin(\pi \nu)  \ln \frac{2 \cal M}{t}  + 2 i \sin (\pi \nu) \frac{ \Gamma ' (\nu+1)}{\Gamma ( \nu +1)}
	\right] \,.
	\end{align}
Note that it is independent of the momentum $\vec{k}$.
Thus the integration for $\vec{k}$ can be done with the coefficient of equation (\ref{eq:fourierQ})
\begin{align}
&\int e^{- i ( k_{1} x_{1} + k_{2} x_{2}) } e^{- \frac{t}{2 \cal M} ( k_{1}^{2} + k_{2}^{2} ) } dk_{1} dk_{2}
= \frac{2\cal M}{t} \pi e^{-\frac{\cal M}{2 t} (x_{1}^{2} + x_{2}^{2})} \,.
\end{align}

As a result, the final form of the inverse Fourier transformation is simply given by
\begin{align}
    \int \frac{d^{2} k }{(2\pi)^{2}}\frac{d\omega}{2\pi} e^{-i \vec k \cdot \vec y + i\omega x^{+}} q^{2\nu} \ln q^{2}
    =-\frac{e^{-\frac{\cal M}{2}\frac{x_{1}^{2}+ x_{2}^{2}}{x^{+}}}}{8 \pi^{2} {\cal M}} \frac{d}{d \nu}
    \left( \Gamma(\nu+1) \sin (\nu \pi)\left( \frac{2 \cal M }{x^{+}} \right)^{\nu+2} \right) \;.
\end{align}
Thus we show that the correlation function $ \langle  \phi (x^+_2, \vec{y}_2) \phi (x^+_1, \vec{y}_1)\rangle $
can be directly computed using the inverse Fourier transformation to confirm the same result obtained by
the $ \nu $ derivative
    \begin{align}
        &\langle  \phi (x^+_2, \vec{y}_2) \phi (x^+_1, \vec{y}_1)\rangle
		= \int \frac{d^2 k}{(2\pi )^2} \frac{d \omega}{2 \pi} e^{i \vec{k} \cdot (\vec{y}_{1} -\vec{y}_{2})
		- i \omega ( x^{+}_{1}-x^{+}_{2} ) } {\cal F}_{2} ( u_{B}, \omega,  \vec{k})    \\
        &\qquad\qquad =  \frac{ \theta( \Delta x^{+}) L^{3}  }{8 \pi^{2} \nu  u_{B}^{4} {\cal M}}
         \exp^{-  \frac{  {\cal M} \Delta \vec{y}^{2} }{2 \Delta x^{+}}  } \cdot
          \frac{d}{d \nu} \left( \frac{\pi \nu}{\Gamma(\nu)} \left(
        \frac{ {2 \cal M}}{ \Delta x^{+}  }\right)^{\nu +2} \left( \frac{u_{B}}{2}\right)^{2\nu} \right)
          \\
        &\qquad \qquad = \frac{1}{2\nu} \frac{d}{d\nu}
        \langle \tilde \phi (x^+_2, \vec{y}_2) \phi (x^+_1, \vec{y}_1)\rangle  \,.
	\end{align}

\end{document}